\pdfoutput=1
\documentclass[journal]{IEEEtran}
\usepackage{cite}
\usepackage{amsmath,amssymb,amsfonts}
\usepackage{algorithmic}
\usepackage{graphicx}
\usepackage{textcomp}

\usepackage{afterpage}
\usepackage{verbatim}
\usepackage{psfrag}
\usepackage{times}
\usepackage{wrapfig}
\usepackage{multirow}
\usepackage{setspace}
\usepackage{epsfig}
\usepackage{epstopdf}
\usepackage{footmisc}
\usepackage{fmtcount}
\usepackage{stackrel}
\usepackage{float}
\usepackage{enumerate}
\usepackage{graphicx}
\usepackage{hyperref}
\usepackage{hypcap}
\usepackage[ruled,vlined,linesnumbered]{algorithm2e}
\usepackage[normalem]{ulem}
\usepackage{authblk}
\usepackage{array}
\usepackage{tabularx}
\usepackage{bbm}

\usepackage{array}
\usepackage{mathtools}
\usepackage[short]{optidef}
\newcolumntype{L}[1]{>{\raggedright\let\newline\\\arraybackslash\hspace{0pt}}m{#1}}
\makeatletter
\newcommand\semiHuge{\@setfontsize\semiHuge{23}{28}}
\makeatother
\usepackage{indentfirst}

\begin{document}

\title{Intelligent Surfaces for 6G Wireless Networks: A Survey of Optimization and Performance Analysis Techniques}

\author{Rawan Alghamdi,
         Reem Alhadrami,
         Dalia Alhothali,
         Heba Almorad,
         Alice Faisal,
         Sara Helal,
         Rahaf Shalabi,
         Rawan Asfour,
         Noofa Hammad,
         Asmaa Shams,
         Nasir Saeed,
         Hayssam Dahrouj,
        Tareq Y. Al-Naffouri,
         and Mohamed-Slim Alouini
 \thanks{The first seven authors contributed equally to the manuscript, and so did the eighth to tenth authors. R. Alghamdi, R. Alhadrami, D.Alhothali, H. Almorad, R. Asfour, A. Faisal, N. Hammad, S. Helal, R. Shalabi, and A. Shams  are with the Department of Electrical Engineering, Effat University, Jeddah 22332, Saudi Arabia. 
 H. Dahrouj is with the Center of Excellence for NEOM Research, Division of Computer, Electrical and Mathematical Sciences and Engineering, King Abdullah University of Science and Technology, Thuwal 23955-6900, Saudi Arabia.
N. Saeed, T. Y. Al-Naffouri, and M.-S. Alouini are with the Division of Computer, Electrical and Mathematical Sciences and Engineering, King Abdullah University of Science and Technology, Thuwal 23955-6900, Saudi Arabia.}}

\maketitle

\begin{abstract}
This paper surveys the optimization frameworks and performance analysis methods for large intelligent surfaces (LIS), which have been emerging as strong candidates to support the sixth-generation wireless physical platforms (6G). Due to their ability to adjust the behavior of interacting electromagnetic (EM) waves through intelligent manipulations of the reflections phase shifts, LIS have shown promising merits at improving the spectral efficiency of wireless networks. In this context, researchers have been recently exploring LIS technology in depth as a means to achieve programmable, virtualized, and distributed wireless network infrastructures. From a system level perspective, LIS have also been proven to be a low-cost, green, sustainable, and energy-efficient solution for 6G systems. This paper provides a unique blend that surveys the principles of operation of LIS, together with their optimization and performance analysis frameworks. The paper first introduces the LIS technology and its physical working principle. Then, it presents various optimization frameworks that aim to optimize specific objectives, namely, maximizing energy efficiency, sum-rate, secrecy-rate, and coverage. The paper afterwards discusses various relevant performance analysis works including capacity analysis, the impact of hardware impairments on capacity, uplink/downlink data rate analysis, and outage probability. The paper further presents the impact of adopting the LIS technology for positioning applications. Finally, we identify numerous exciting open challenges for LIS-aided 6G wireless networks, including resource allocation problems, hybrid radio frequency/visible light communication (RF-VLC) systems, health considerations, and localization.
\end{abstract}
\IEEEpeerreviewmaketitle


\section{Introduction}
\label{sec:introduction}
The recent advent of large intelligent surfaces (LIS) empowers smart radio environments at overcoming the large power consumption and the probabilistic nature of electromagnetic (EM) wave transmission, thereby improving both the quality of service (QoS) and radio connectivity \cite{basar2019wireless}. In  other  words,  LIS  combat  the  complex  phenomena  of EM  propagation coupled  with  parameters  that  cannot  be measured precisely by means of tactfully controlling the channel based on the wireless network geometry. Therefore, LIS are envisioned to be one of the essential technology enablers of 6G and beyond wireless communications\cite{yuan2020reconfigurableintelligentsurface}. By means of combating the uncontrollable and stochastic wireless propagation medium, LIS realize a controllable and smart radio environment in a software-controlled fashion, which boosts the communication capabilities. Software-defined or reconfigurable EM meta-surfaces are the fundamental technology behind the LIS implementation that is capable of modulating data onto the received signals\cite{basar2019reconfigurable}, customizing changes to the radio waves, and intelligently sensing the environment. In other words, intelligent meta-surfaces are programmable frequency-selective surfaces that are composed of artificial thin meta-material films, that are adequate for energy-efficient and low-complexity wireless communications \cite{Hu_datatx, energEff}.  The classical utilization of LIS technology was initially restricted to satellite and radar communication systems, and was not adopted by terrestrial wireless communications. The conventional wireless radio transmission rather relies on traditional reflecting surfaces, which only induce fixed phase shifts, and do not adapt to the terrestrial time-varying wireless communication channels. Fortunately, the recent advances in metamaterials and micro-electro-mechanical systems (MEMS) lead to the advance of reconfigurable reflecting surfaces. For instance, Kaina \textit{et al.}  first introduced the concept of LIS by using tunable surfaces to control the wireless propagation environment\cite{kaina2014shaping}. In this regard, LIS-assisted communication can be seen as an enhanced platform of  conventional wireless communication systems, as LIS bring in more degrees of freedom via controlling the wireless channel. Hence, optimization of the wireless channel leads to a more relaxed set of constraints, yielding an increase in the overall system performance.

Moreover, meta-surfaces are immune to noise in radio receivers and do not require either analog/digital converters, or power amplifiers. Meta-surfaces can, therefore, manipulate and reflect the signals with extremely low-noise amplification. Enabling the LIS technology would also allow for reductions in power consumption as compared to current wireless networks. Due to its relative high energy-efficiency, LIS technology is environmentally friendly, as its reduces the overall carbon footprint \cite{basar2019wireless,8644519Huang}.

From a spectrum deployment perspective, recent studies show that high-frequency transmission, such as millimeter wave (mmWave) and terahertz (THz) communications, can be well realized using LIS \cite{MA2020}. Providentially, significant disadvantages for the use of LIS are not known at the moment for very high frequency deployments, especially at the range of 2.4 GHz to 60 GHz. This attractive attribute of LIS makes them suitable for use in cutting edge communication and sensing technologies of 5G and beyond systems, which include Internet of Things (IoT), Device-to-Device (D2D) communications, Machine-to-Machine (M2M) communications, etc. \cite{liaskos2018using,renzo2019smart, 8449754Liaskos}.

The above unique features of LIS make them suitable for a variety of groundbreaking smart radio environments applications, both in indoor and in outdoor environments \cite{Beam_Wu}. For instance, in \cite{6231145Subrt}, the authors introduce the deployment of a smart indoor environment using the concept of intelligent walls empowered by machine-learning control algorithms for realizing an indoor cognitive wireless network. In indoor environments, various objects including walls, furniture, and windows, can influence the communication and coverage for wireless devices. As indoor scenarios (e.g., hospitals, hotels, security offices) require ultra-reliable high-speed communications, the authors in \cite{8449754Liaskos} propose coating the indoor objects with software-programmable hyper-surface tiles (a novel class of meta-surfaces), so as to improve both communication and coverage aspects of indoor wireless systems. LIS have indeed many other prospective applications, including indoor localization \cite{positionmmWave}, health monitoring using smart T-shirts \cite{renzo2019smart}, imaging, quantum optics, and military purposes \cite{8449754Liaskos}.

Thanks to the converging breakthrough in developing smart surfaces, boosting the operation of smart radio environments with intelligent and reconfigurable capabilities is more feasible than ever. Such emerging capabilities would indeed enable the network operators to shape the radio waves propagation with customized functionalities. For example, the embedding of meta-surfaces into the outdoor and indoor objects would allow for sensing the incoming signal response and feeding the system response back to the network controller \cite{Beam_Wu}. Based on the sensed data, meta-surfaces can then configure and manipulate the input signal wave through different EM behavior control functions, e.g., wave reflection, refraction, polarization, or full absorption. Moreover, reflection and refraction functions can offer additional services, also known as \textit{wave steering}, that can override the conventional Snell's law\cite{8449754Liaskos}. Through extending the notion of network softwarization, a smart radio environment can facilitate programmatic commands, and can be remotely configured and/or elastically optimized. Without generating new signals, which consume an additional extent of power, LIS are, therefore, able to meet the challenging requirements of future wireless networks \cite{liaskos2018using}.

It is worthy to note that LIS are complementary mediums to support other emerging technologies, including backscatter communication, millimeter-wave communication, massive multiple-input multiple-output (mMIMO), and network densification. For instance, LIS, albeit being distinct from mMIMO, can be viewed as an extension of conventional MIMO systems. In \cite{zhao2019survey}, the authors compare mMIMO with Intelligent Reflecting Surfaces (IRS) deployment in terms of information transfer capabilities. Although mMIMO  boosts the energy and spectral efficiencies of the communication links, it is not capable of tuning and controlling the wireless propagation environment. Moreover, unlike the case of IRS, for which the capacity is linearly proportional with the average transmit power \cite{zhao2019survey}, a logarithmic relationship exists between the capacity and average transmit power in mMIMO. However, in a matching array architecture, mMIMO setups can achieve higher signal-to-noise ratio (SNR) with reduced power consumption than in IRS-aided setups operating in a far-field region \cite{bjrnson2020power}.

Furthermore, the implementation of the IRS-assisted non-orthogonal multiple access system (NOMA) has drawn significant attention for optimizing the performance (connectivity and spectral efficiency) of the next generation multi-user wireless communication networks. Recent works in \cite{hou2020mimonoma,cheng2020downlink,YueNOMA,zuo2020resource,hou2019reconfigurable,9000593Ding,ding2020impact} investigated the potential applications of integrating NOMA technology with IRS-based communications.

 Additionally, incorporating full-duplex (FD) communication into IRS can exploit new degrees of freedom, facilitating spectrum-efficient and low-cost designs for next-generation wireless communication systems. Also, FD-enabled IRS systems can be further employed for other applications, e.g., cooperative jamming \cite{pan2020fullduplex}.

\begin{table*}[ht]
\centering
\caption{ A list of LIS surveys.} \label{tab:par}
\begin{tabular}{|p{0.7cm}|p{7cm}|p{8cm}|c|}
    \hline
    \textbf{Ref.}  & \textbf{Focus} &\textbf{Comparison to our paper} \\
    \hline
     \cite{basar2019wireless}  & Outlines different types of reconfigurable surfaces and their theoretical performance limits  & \multirow{12}{8cm}{Unlike these works, our paper surveys various critical and technical aspects of LIS, including optimization frameworks and performance analysis. The paper entails the physical working principle of LIS. It then introduces the optimization schemes for LIS-based systems, which include energy efficiency, sum-rate and secrecy-rate. The paper afterwards surveys various performance analysis metrics including capacity analysis, uplink/downlink data rate analysis, and outage probability. Lastly, we suggest various open research issues on the LIS topic.}  \\ \cline{1-2}
    \cite{6648436S}  & Explores the capabilities of reconfigurable reflectarrays for multi-band operation, amplification, and polarization manipulation  & \\  \cline{1-2}
       \cite{renzo2019smart}  & Motivates for using LIS in future wireless communication networks, especially for reducing power consumption and improving connectivity &\\ \cline{1-2}
     \cite{zhao2019survey}  & Summarizes studies on intelligent reflecting surfaces, mainly their applications & \\  \cline{1-2}
    \cite{gong2019smart}  & Focuses on design and applications aspects of intelligent reflecting surfaces & \\
     \hline
\end{tabular}
\end{table*}

 \subsection{Related Surveys}
The topic of intelligent surfaces has rapidly attracted research interests with preliminary contributions to communication-theoretical modeling, optimization, deployment, and design of LIS-empowered networks.
Hence, some recent surveys overview the revolutionary technology of LIS and their promising significance in future wireless communication networks. In \cite{basar2019wireless}, the authors outlined the state-of-the-art reconfigurable surfaces solutions, 6G applications, and theoretical performance limits. Recently, Jun \textit{et al.} provided a general overview of the data rate and reliability issues in LIS \cite{zhao2019survey}. The use of LIS in network security, channel estimation, and deep learning-based paradigm for LIS-aided communications were also discussed in \cite{zhao2019survey}. In the context of practical implementation of LIS technology, Sean \textit{et al.} surveyed the primary design architectures, such as the reconfigurable-array lens and reflect-array antennas \cite{6648436S}. Furthermore, the authors in \cite{renzo2019smart} presented the recent research efforts to deploy smart radio environments in practice, which is a step forward towards redefining the current network communication models. The authors in \cite{gong2019smart} then provided an overview of the performance analysis and optimization in LIS-assisted networks, as a means to achieve different wireless communications objectives. In \cite{gong2019smart}, the authors further discussed the diverse application scenarios that can exploit LIS, such as wireless power transfer, mobile edge computing, and unmanned aerial vehicle (UAV) based communication. Table \ref{tab:par} provides an illustrative summary of the above recent LIS surveys.

Unlike all existing works, the current survey is the first of its kind which overviews the technical and critical aspects of mathematical optimization and performance analysis of LIS systems, and presents a handful of promising research directions towards the formulations of practical problems in future beyond 5G systems. More specifically, our paper first entails the physical working principle of LIS. It then introduces the optimization schemes for LIS-based systems, which include energy efficiency, power optimization, sum-rate, secrecy-rate, and coverage. The paper afterwards surveys various performance analysis works including those which tackle capacity analysis, the impact of hardware impairments on capacity, uplink/downlink data rate analysis, and outage probability. The paper also presents the localization error performance of centralized and distributed LIS systems. Lastly, we suggest various open research issues in the context of future LIS-empowered systems.
\vspace{-1.0em}
\subsection{Organization of the Survey} 
The rest of the paper is organized as follows. Section \ref{sec:workingPrinc} discusses the working principle of LIS, providing the associated enabling technologies and models for the control of EM wave reflection. Section \ref{opt} covers the optimization techniques for LIS, while Section \ref{sec:perf} sheds light on the performance analysis of LIS. Section \ref{sec:RelAnalysis} discusses the error performance and reliability analysis of LIS. In section \ref{sec: position}, recent studies on the potential of positioning and coverage in LIS systems are surveyed. Before concluding in section \ref{sec:conc}, various future research directions are discussed in section \ref{sec:openresearch}. Finally, Table \ref{tab:acro} lists the mostly utilized acronyms in the manuscript for the convenience of the readers. 

\begin{table}[ht]
\centering
\caption{List of mostly used acronyms} \label{tab:acro}
\begin{tabular}{|p{0.09\textwidth}|p{0.34\textwidth}|}
    \hline
    \textbf{Acronym} & \textbf{Definition} \\ \hline\hline
     AP & Access-Point \\ \hline
     BS & Base station \\ \hline
     CRLB & Cram\'{e}r-Rao Lower Bounds \\ \hline
     CSI & Channel State Information \\ \hline
     EE & Energy Efficiency\\ \hline
     EM & Electromagnetic \\ \hline
     EHR &  Energy Harvesting Receiver\\ \hline
     HWI &  Hardware Impairment\\ \hline
     IDR &  Information Decoding Receiver \\ \hline
     IRS & Intelligent Reflecting Surfaces \\ \hline
     LIS & Large Intelligent Surfaces \\ \hline
     LoS & Line-of-Sight \\ \hline
     MEMS & Micro-electro-mechanical Systems \\ \hline
     MIMO & Multiple input multiple output \\ \hline
     MISO & Multiple input single output \\ \hline
     mMIMO & massive Multiple input multiple output  \\ \hline
     mmWave & millimeter wave \\ \hline
     OLP & Optimal Linear Precoder \\ \hline
     QoS & Quality-of-Service \\ \hline
     RIS & Reflecting Intelligent Surfaces \\ \hline
     RSS & Received Signal Strength\\ \hline
     SINR & Signal-to-interference-plus-noise ratio\\ \hline
     SNR & Signal-to-noise ratio \\ \hline
     SSE & System Spectral Efficiency \\ \hline
     SWIPT & Simultaneous Wireless Information and Power Transfer \\ \hline
     UE & User Equipment\\ \hline
     VLC & Visible Light Communication\\ \hline
\end{tabular}
\end{table}

\section{Working Principle}\label{sec:workingPrinc}
This section sheds light on the basic working principle of LIS from a physical layer perspective. We first note that there are several terminologies that are interchangeably used to denote LIS in the literature, namely, IRS, reconfigurable intelligent surfaces (RIS), and software-defined surfaces (SDS). Each of such terminologies emphasizes one particular feature of the smart surface \cite{zhao2019survey}. For completeness, we refer to Table \ref{table:LIS} for a comparison of such various terms.

LIS consists of a two-dimensional array made up of numerous low-cost nearly passive reflecting EM elements (except for few active elements connected to the LIS controllers) \cite{najafi2020physics}.  Nevertheless, some studies show the potential of active element-based LIS designs \cite{Hu, jung_datarate, SSE}. In fact,  the latest  LIS-aided hybrid wireless networks,  comprising of both active and passive elements, are projected to be promising to achieve cost-effective sustained performance development \cite{wu2020intelligent}. In our paper, we focus on the working principle of the passive elements, also known as metamaterials. Such metamaterials consist of either varactor diodes or other MEMS that can intelligently adjust their induced phase shifts to attain the desired communication objectives \cite{nadeem_asymptotic}. They consist of repeated meta-atoms over a substrate with a specific EM behavior. The metamaterial's  EM behavior depends on the meta-atom structure. Hence, some patterns of meta-atoms absorb the entire incoming EM waves, while other models may entirely reflect the incoming EM waves. The metasurfaces are dynamic, consisting of tunable elements that can switch their condition and  EM behavior by applying an external bias. These tunable elements may include CMOS switches or MEMS switches. In the metasurfaces, the switching elements control the meta-atoms that act as input and output antennas. Besides, the switching elements also connect the meta-atoms in custom topologies. That is, when incoming EM waves enter from an input antenna, they are routed based on the status of the switch, and exit via the output antenna, helping the LIS to achieve a customized reflection \cite{8449754Liaskos}.

\begin{figure}
    \centering
    \includegraphics[width=0.7\linewidth]{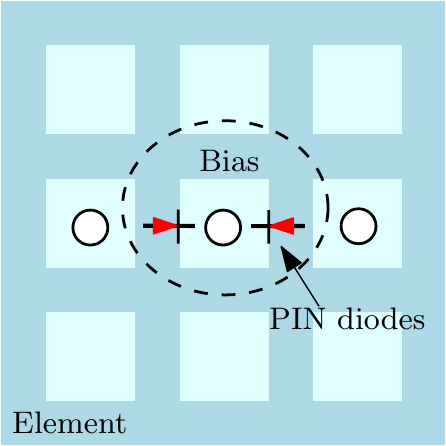}
    \caption{Controlling the EM reflection using PIN diodes.}
\end{figure}

There are various switching technologies to control the EM reflection from the smart surface, including positive-intrinsic-negative (PIN) diodes, varactor-tuned resonators, liquid crystal, and MEMS technologies. One way of controlling the reflection effect in a metasurface is by placing PIN diodes as switch elements.
An external bias switches the PIN diodes on and off, generating two different states for the smart surface, as shown in Fig.1. When the PIN diode is turned off, the incoming energy penetrates the surface and is mostly absorbed. However, when the PIN diode is on, most of the incoming energy is reflected  \cite{basar2019wireless}.
\begin{figure}
    \centering
    \includegraphics[width=\linewidth]{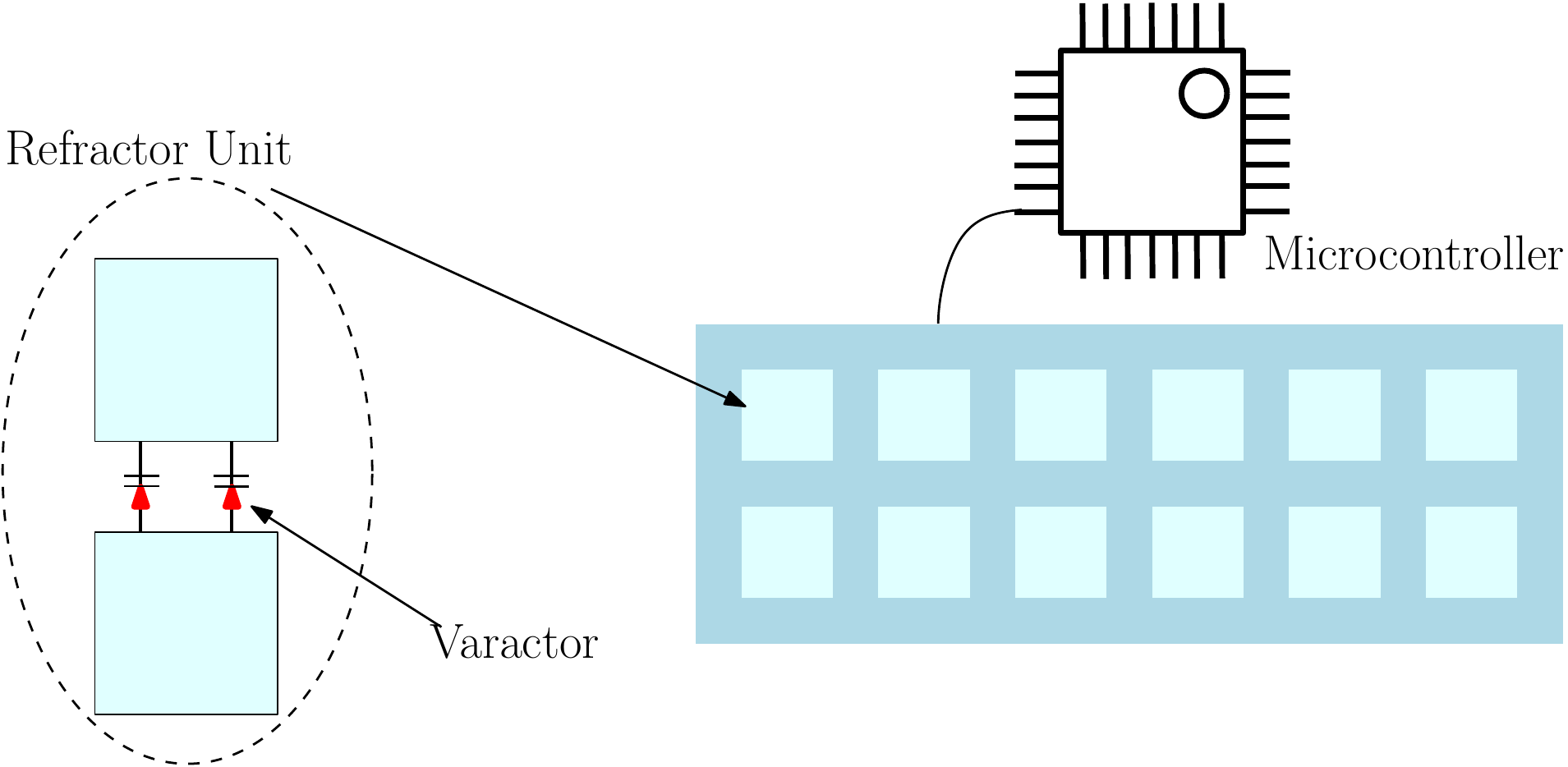}
    \caption{Controlling the EM reflection using varactor-tuned resonators. \label{fig:vractor}}
    \end{figure}

Additionally,  varactor-tuned resonators are also used for controlling the signal's propagation, as illustrated in Fig. 2. When the bias voltage is applied to the varactor, a tunable phase shift is attained. The liquid crystals can further tune the phase shift of the reflected signal, as suggested in \cite{8073077}. By differing the direct current (DC) voltages on the patches of liquid crystal-loaded unit cells, the effective dielectric constant of any individual unit can be thus adjusted. As a result, the phase shifts of the incoming signal can be controlled at various locations of the metasurface. Dynamic metasurfaces make up a tile that consists of a gateway, to which the controller network acquires a slave/master relationship.  The controller network records its running state and receives instructions to change the current condition of the switching elements through the gateway \cite{8449754Liaskos}.

\begin{table*}[ht]
\centering
\caption{List of different LIS terms.}\label{table:LIS}
\begin{tabular}{|p{3.7cm}|p{5.0cm}|p{8.0cm}|}
 \hline
    \textbf{Ref.}                                                           & \textbf{Surface Term}                                 & \textbf{Reason} \\
       \hline
\cite{Hu_datatx},\cite{jung_datarate},\cite{8108330Hu}    &  Large Intelligent Surfaces (LIS)  & Considers limitless surface length or a massive number of antennas.    \\ \hline
\cite{zhao2019survey}, \cite{unknownshi},\cite{Wu_BeamOpt},\cite{8910627}   &  Intelligent Reflecting Surfaces (IRS)       & Emphasizes more on the reflecting property of the smart surface.           \\ \hline
\cite{basar2019wireless},\cite{energEff}     &  Reconfigurable Intelligent Surfaces (RIS)   &  Highlights more the reconfigurability of the smart surface for the incident signal.                     \\ \hline
\cite{8683663},\cite{Huang_rate}                 &  Passive Intelligent Surfaces (PIS)          & Underlines the passive reflection with no power consumption.   \\ \hline
\cite{renzo2019reflection}, \cite{spacetime}           &  Reconfigurable Metasurfaces (RM)            &  Emphasizes more on the metalic pattern through which the surface is engineered.                   \\ \hline
\cite{basar2019reconfigurable}              &  Software-Defined Surfaces (SDS)             & Considers the software-defined interaction between the surface and incoming waves.                  \\ \hline
\cite{jung_datarate}             &  Software-Defined Metasurfaces (SDM)         &  Emphasizes on both the metallic pattern and the software-defined function.                  \\ \hline
\cite{8879620He}                        &  Large Intelligent Metasurfaces (LIM)        & Assumes massive number of antennas for the asymptotic analysis of metasurfaces.                \\ \hline
\cite{8485924}, \cite{7510962}              &  Smart Reflect-Arrays (SRA)                       & Pinpoints the reflection function.        \\
 \hline
\end{tabular}
\end{table*}

\begin{figure}
    \centering
    \includegraphics[width=\linewidth]{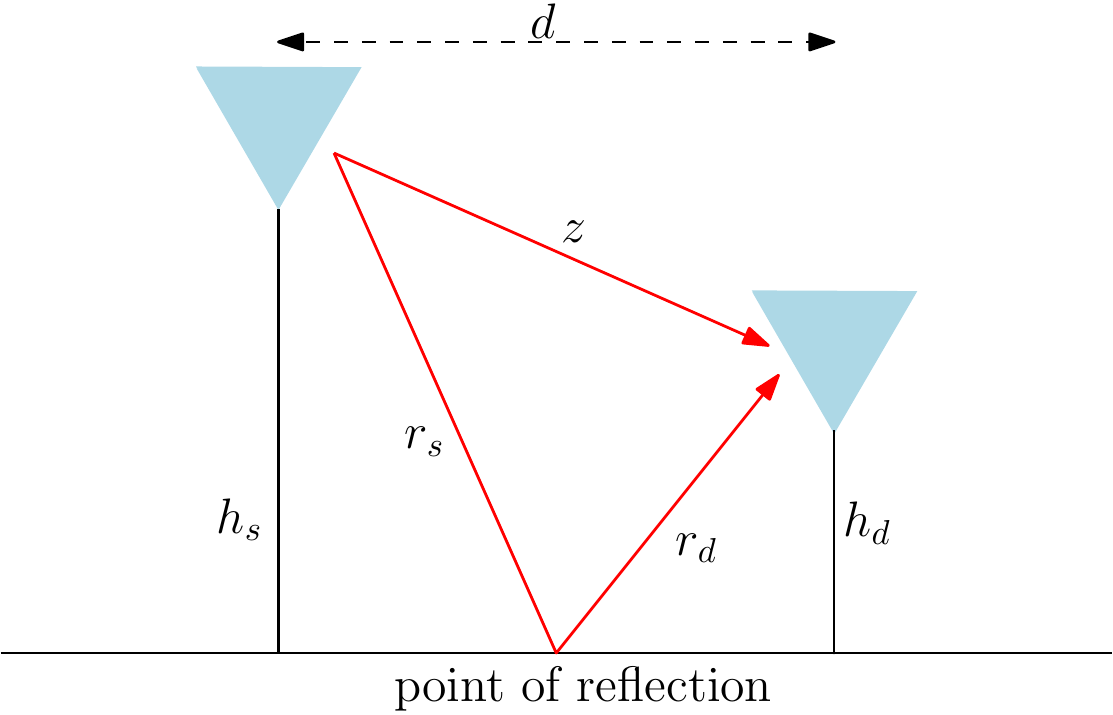}
    \caption{Conventional two-ray propagation model. \label{fig:LIS}}
    \end{figure}

To get a better understanding of such working principle, we next demonstrate a basic example of a controllable wireless propagation by means of inducing an intelligent surface. The example considers the conventional two-ray channel model for a free space environment and a reflecting surface deployed on the ground plane  \cite{basar2019wireless}.

The propagation of radio waves are described in terms of rays using a ray optics model which assumes that the geometric size of the ground plane is considerably larger than the wavelength of the radio wave and that the ground plane reflections are specular \cite{born_wolf_1997}. In addition, the model proposes that the radio waves travel in straight lines in case of homogeneous media, i.e., the energy is transported along certain curves. More concisely, the model adheres to Fermat's principle that states that the ray travels along the path between two points with minimum travelling time. The received signal is composed of the line-of-sight (LoS) ray and the reflected ray from the ground, as depicted in Fig. \ref{fig:LIS}. Then, according to Snell's law of reflection, the point of reflection where the imaginary vertical line stands. The angle between the incident ray and this vertical line is equal to the angle between the reflected ray and the imaginary line. Based on this model, the power received at the destination is represented as
\begin{equation}
\label{eq: RS}
P_d = P_s \left(\frac{\lambda}{4\pi}\right)^2 {\vline\frac{1}{z}+\frac{R \times e^{-j\Delta\phi}}{r_s+r_d}\vline~}^2
\end{equation}
where $P_s$ is the power transmitted from the source, $\lambda$ is the wavelength, $z$ is the distance between the source and the destination antennas, $r_s$ is the distance between the source antenna and the point of reflection on the LIS, $r_d$ is the distance between the point of reflection and the destination antenna, $R$ is the ground reflection coefficient, and $\Delta \phi = \frac{2 \pi (r_s +r_d-z)}{\lambda}$ represents the phase difference between the two paths.
If we assume that there is no ground reflection, and the distance between the source and the destination is very large, i.e., $d >> h_s + h_d$, then $d \approx z \approx r_s+r_d$, and therefore (\ref{eq: RS}) can be simplified as
\begin{equation}
\label{eq: RS2}
P_d = P_s \left(\frac{\lambda}{4\pi d}\right)^2.
\end{equation}
Both \eqref{eq: RS} and \eqref{eq: RS2} show that the uncontrollable reflection from the ground surface degrades the received power of the signal.

\begin{figure}
    \centering
    \includegraphics[width=\linewidth]{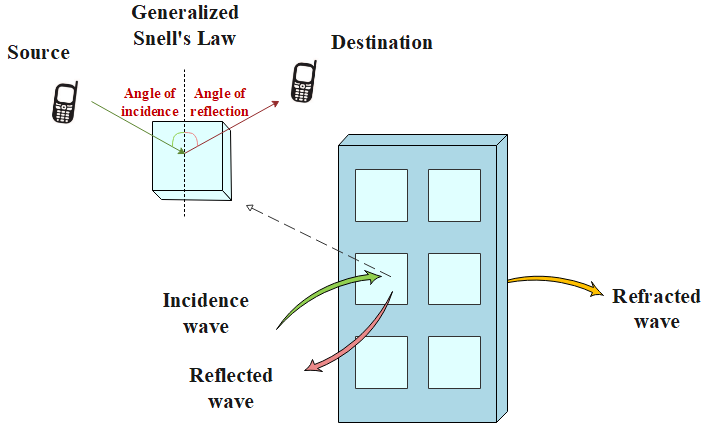}
    \caption{The working principle of LIS as a reflector. \label{refl}}
    \end{figure}
Consider an LIS system with $N$ reconfigurable meta-surfaces laid on the ground, and acting as a  reflecting surface to support the communication between the source and the destination, as depicted in Fig. 4. According to the generalized Snell's law, each meta-surface can independently tune the angle of reflection and the reflected ray phase. Moreover, unlike a typical passive surface, the incidence and reflection angles are not necessarily the same. Thus, the power received at the destination with respect to $i$-th reconfigurable meta-surface can be represented as 

\begin{equation}
\label{eq: RS1}
P_d = P_s \left(\frac{\lambda}{4\pi}\right)^2 {\vline\frac{1}{z}+\sum_{i=1}^{N}\frac{R_i \times e^{-j\Delta\phi_i}}{r_{s,i}+r_{d,i}}\vline~}^2.
\end{equation}
To avoid the fluctuation of the power received at the destination, the reflection coefficients of the reconfigurable meta-surfaces are optimized based on various assumptions, such as the absence of reflection losses and perfect LIS phase knowledge. Hence, every $R_i$ is optimized to align the phase of the received signal with the LoS path. Then, the power received at the destination can simply be written as follows:
\begin{equation}
P_d \approx (N+1)^2 P_s \left(\frac{\lambda}{4\pi d}\right)^2. 
\end{equation}
In conclusion, the power received at the destination is directly proportional to the square of the number of the independently controlled phases of LIS $N^2$ and inversely proportional to the square of the distance between the source and the destination. This clearly shows the potential of LIS in wireless networks where a power gain is obtained as a function of the number of reflecting meta-surfaces \cite{basar2019wireless}.

To best realize the full potential of LIS systems, our paper next focuses on ways of optimizing the performance of LIS systems through surveying the major relevant optimization frameworks. The paper further addresses the performance analysis of LIS systems, and illustrates several open issues in the context of LIS-aided wireless networks.

\section{Optimization Use Cases in IRS-based Systems}
\label{opt}

\begin{figure}
    \centering
    \includegraphics[width =0.48\textwidth]{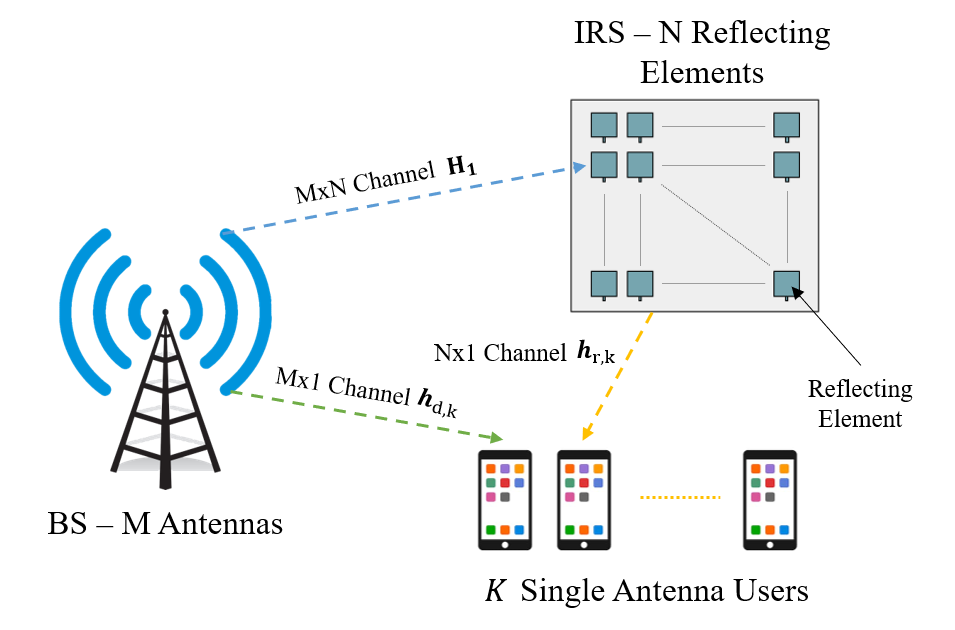}
    \caption{An IRS-aided wireless communication system from a BS. \label{fig: basic}}
    \end{figure}

IRS are proposed as intrinsic components of beyond-5G wireless systems, as they have the potential of transmitting data through multiple active elements, and intelligently adjusting the communication channel in the process \cite{Jensen2020}. The increasing demands for data rate requirements and higher-speed wireless communications for future networks have raised serious concerns on their power consumption, energy efficiency, secrecy rate, etc. As discussed earlier, IRS are considered as contiguous surfaces of electromagnetically active materials. Thus, to realize the full potential of IRS systems, they have to be well-designed, optimized and integrated. Hence, in this section, we survey the optimization frameworks of IRS, including the maximization of energy efficiency, sum-rate, secrecy-rate and coverage. Fig. \ref{fig: basic} illustrates the basic structure of IRS system, where $N$ indicates the number of reflecting elements in the IRS, and $M$ indicates the number of transmit antennas at the base station (BS). IRS are connected to $K$ single-antenna users, where $\boldsymbol{h}_{d,k}$, $\boldsymbol{h}_{r,k}$, and $\boldsymbol{H}_1$ denote the channel links between the IRS to the $k$-th user, the BS to the $k$-th user, and the BS to the IRS, respectively. Table \ref{tab:par2} contains the list of symbols that are used in the following sections.

\begin{table}[ht]
\centering
\caption{List of Symbols} \label{tab:par2}
\begin{tabular}{|c|c|}
    \hline
    \textbf{Symbol} & \textbf{Definition} \\ \hline\hline
     $\boldsymbol{\Phi} $ & Phase shift vector \\ \hline
     $\boldsymbol{\Theta}$ & Diagonal matrix of the effective phase shifts \\ \hline
     $P_{max}$ & Maximum transmit power \\ \hline
     $\boldsymbol{h}_r$ & Channel between IRS-user \\ \hline
     $\boldsymbol{h}_d$ & Channel between AP-user\\ \hline
     $\boldsymbol{G}$ & Channel between AP-IRS \\ \hline
     $\boldsymbol{H}_1$ & Channel between BS-IRS \\ \hline
     $\boldsymbol{h}_{r,k}$ & Channel between IRS-user $k$\\ \hline
     $\boldsymbol{h}_{d,k}$ & 
     Channel between BS-user $k$ \\ \hline
     $\boldsymbol{w}$ & Beamforming vector \\ \hline
     $p_k$ & Transmit power \\ \hline
     $P_{BS}$ & BS's total power consumption \\ \hline
     $P_{UE}$ & Hardware static power \\ \hline
     $P_n(b)$ & Phase shifter's power consumption \\ \hline
     $\boldsymbol{w}_i$ & Precoding vectors for IDR\\ \hline
     $\boldsymbol{v}_j$ & Precoding vectors for EHR \\ \hline
     $\boldsymbol{S}$ & Trace of a positive semi-definite matrix \\ \hline
     $K$ & Number of sub-carriers \\ \hline
     $\boldsymbol{v}$ & IRS reflect beamforming vector \\ \hline
    $\boldsymbol{W}$ &Transmit beamforming matrix at the BS \\ \hline
    $\boldsymbol{x}$ & Transmit signal\\ \hline
    $\gamma_k$ & Downlink SINR at the $k$th user\\ \hline
$\sigma^2$ & Noise variance
\\ \hline
\end{tabular}
\end{table}

\subsection{Energy Efficiency}
With the rapid growth of wireless networks, the number of connected devices continues to increase exponentially, leading to dense deployment of MIMO base-stations and access-points (APs). Since IRS comprise a massive number of reflecting elements, their deployments would enhance the efficiency of future wireless networks, as high passive beamforming gains can be collaboratively achieved via modifying the phase shifts of the reflected signals \cite{Nie2020}.

As energy efficiency (EE) is a crucial performance metric for balancing the throughput and the power consumption, the question of reaching an optimized strategy to reach a maximal EE performance is of high importance in IRS. To this end, the work in \cite{energEff} proposes a significant sustainable energy-efficient approach. Since IRS-based systems are capable of adjusting the phase shift induced by each reflecting element to constructively combine the reflected signal, they can amplify and forward the signals without the need for additional power amplifiers. To this end, IRS-based systems become more favorable than conventional amplify-and-forward (AF) relay systems \cite{AF}. In particular, an optimized EE policy in IRS systems via optimizing the phase shifts and transmit power is proposed in \cite{energEff}, while satisfying specific power and QoS constraints \cite{energEff}.
More precisely, in \cite{energEff}, the authors consider an IRS-based downlink multi-user multiple-input-single-output (MISO) system with $K$ users, and one BS equipped with $M$ antennas as shown in Fig.\ref{fig: basic}. Following the analysis of \cite{energEff}, the EE maximization problem is addressed as:
\begin{maxi!}
		{\boldsymbol{\Theta},\boldsymbol{P}} {\frac{\sum_{k=1}^K \log_2(1+p_k \sigma^{-2}) }{\xi \sum_{k=1}^K p_k + P_{BS} + KP_{UE} + N P_n (b)}}
		{\label{eq:ee}}{}
		\addConstraint{\log_2(1+p_k \sigma^{-2}) \geq R_{\text{min},k},\quad \forall k = 1,2,...,K} \label{eq:fs}
		\addConstraint{\text{tr}((\boldsymbol{H}_2 \Theta \boldsymbol{H}_1)^+ \boldsymbol{P} (\boldsymbol{H}_2 \Theta \boldsymbol{H}_1)^{+\dagger}}) \leq P_{max}\label{eq:sn}
		\addConstraint{|\Phi_n|=1 ~\forall n=1,2...,N}\label{eq:th}
	\end{maxi!}\label{1}
	
\noindent where $\xi = \eta^{-1}$, $\eta$ denotes the efficiency of the power amplifier at the transmitter side,  $\boldsymbol{H}_1$ is the channel matrix
between the BS and the IRS, $\boldsymbol{H}_2$ is the compound channel matrix between the IRS and all users, i.e., $\boldsymbol{H}_2$ = $[\boldsymbol{h}^T_{r,1} \hspace{2pt} \boldsymbol{h}^T_{r,2} \hspace{2pt} ... \hspace{2pt} \boldsymbol{h}^T_{r,K}]^T$, where $\boldsymbol{h}^T_{r,k}$  denotes the channel vector between the IRS and user $k$, and $p_k, P_{BS}, P_{UE}, P_n(b)$ denote the transmit power of user $k$, the total hardware power consumption at the BS, the hardware static power, and the power consumption of each phase shifter for $b$ number of bits, respectively. Moreover, the total signal power is denoted by $\boldsymbol{P}=\text{diag}(p_1,...., p_k)$.
The constraint in (\ref{eq:fs}) accounts for the individual QoS requirement, i.e., $R_{\text{min},k}$ of the $k$-th user. Constraint (\ref{eq:sn}) denotes the power budget, where $\boldsymbol{H}_2 \Theta \boldsymbol{H}_1$ is the equivalent channel matrix, $\boldsymbol{\Theta}$ is the diagonal matrix that accounts for the effective phase shifts of the IRS elements, $\text{tr}(\cdot)$ is the trace operator, and $P_{max}$ is the maximum power transmitted by the BS. Note that the superscripts $\dagger$ and $+$ indicate the Hermitian (conjugate transpose) and the pseudo-inverse of a matrix, respectively. The constraint in (\ref{eq:th}) discretizes the IRS phase shifts. The above formulated problem in (\ref{eq:ee}) is a non-convex optimization problem. To this end, the work in \cite{energEff} develops two efficient approaches to solve problem (\ref{eq:ee}). Firstly, an alternating optimization technique is employed that iteratively solves for both $\boldsymbol{\Theta}$ and $\boldsymbol{P}$. Secondly, both a gradient descent and a sequential fractional programming (SFP) are adopted to solve the problem. The results in \cite{energEff} illustrate that an optimized IRS-based system achieves an EE gain in the order of 300\% as compared to the conventional AF-based systems.

\begin{figure}
    \centering
    \includegraphics[width =0.48\textwidth]{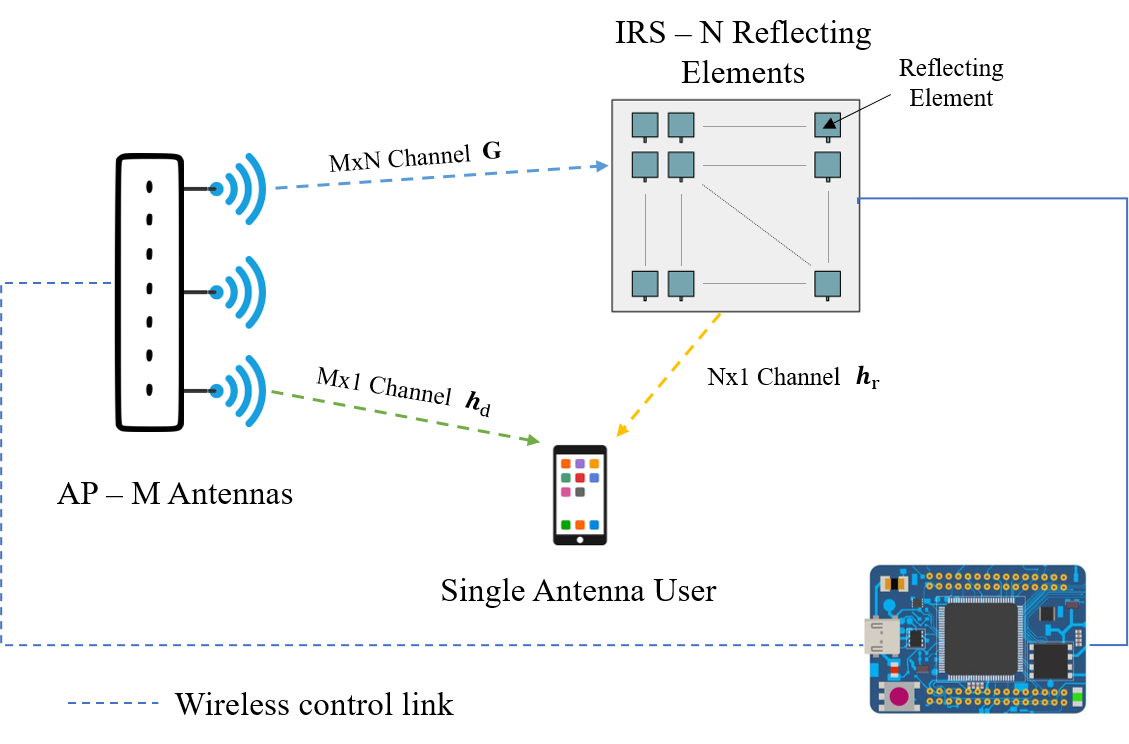}
    \caption{An IRS-aided wireless communication system from an AP. \label{fig: APbasic}}
    \end{figure}

\begin{figure}
    \centering
    \includegraphics[width =0.48\textwidth]{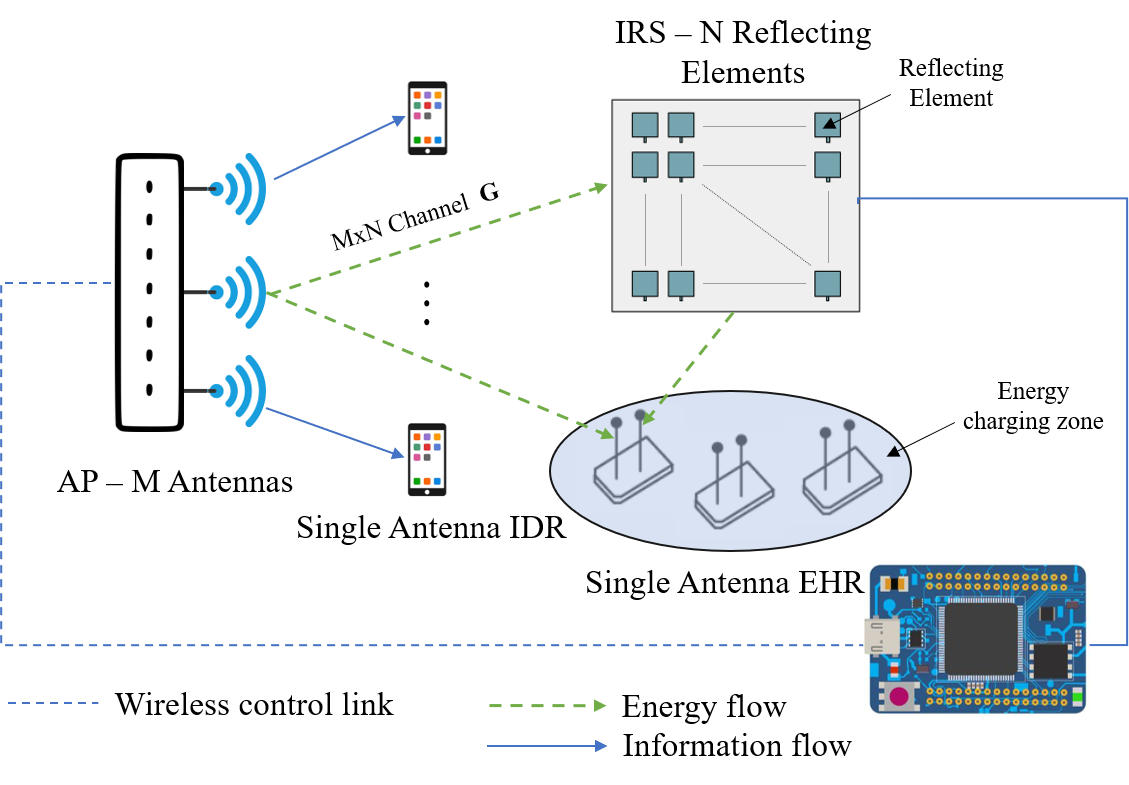}
    \caption{An IRS-aided SWIPT system. \label{fig: APSWIPT}}
    \end{figure}

\subsection{Power Optimization}

In addition to EE, power optimization frameworks are prerequisites for efficient utilization of future wireless networks. The majority of existing works on IRS assume continuous phase shifts for all reflecting elements, which is not practical due to  hardware limitations \cite{Beam_Wu, Huang_rate}. In \cite{Wu_BeamOpt},  Wu \textit{et al.} considered an IRS-assisted system which employs discrete phase shifts at each element to support the communication between the AP and the single antenna user. Configuring the phase shifts results in adding the reflected and non-reflected signals together constructively by the IRS, thereby improving the desired signal power and the wireless network performance. Note that the interference-free zone is created by the IRS due to their spatial interference cancellation capability, which holds only by assuming an ideal case, i.e., where the phase shift at each reflecting element is continuous. This assumption, however, is not practical; therefore, the work in \cite{Wu_BeamOpt} considers a more practical approach with a limited number of discrete phase shifts, but with continuous transmit beamforming vectors at the AP.

The system model in \cite{Wu_BeamOpt} is a conventional MISO setup, consisting of $N$ reflecting elements, a receiver with a single antenna, and an AP equipped with $M$ antennas as shown in Fig. \ref{fig: APbasic}. This model is developed for downlink communication assuming a quasi-static flat-fading channel. Moreover, \cite{Wu_BeamOpt} only accounts for the signals that are reflected by the IRS once. The objective in \cite{Wu_BeamOpt} is to minimize the transmitted power at the AP by optimizing the transmit beamforming and passive reflect beamforming vectors at the AP and at the IRS, respectively, while satisfying a signal-to-noise-ratio (SNR) threshold at the receiver side. The problem in \cite{Wu_BeamOpt} is formulated as follows:

\begin{mini!}
		{\boldsymbol{w},\boldsymbol{\Phi}}    {\| \boldsymbol{w}\|^2 }
		{\label{eq:minp}}{}
		\addConstraint{|(\boldsymbol{h}_r^\dagger \boldsymbol{\Theta} \boldsymbol{G} + \boldsymbol{h}_d^\dagger)}\boldsymbol{w}| \geq \gamma \sigma^2 \label{eq:snd}
		\addConstraint{\Phi_n \in \mathcal{F}, \forall n}\label{eq:pha}
	\end{mini!}

\noindent where $\|\boldsymbol{w}\|^2$ denotes the total transmit power, $\boldsymbol{\Phi}$ is the phase shift vector $\boldsymbol{\Phi}=[\Phi_1,...,\Phi_N] $, $\boldsymbol{h}_r^\dagger$, $\boldsymbol{h}_d^\dagger$, and $\boldsymbol{G}$ represent the channel links between the IRS to user, the AP to user, and the AP to IRS, respectively. The constraint (\ref{eq:snd}) assures that the SNR at the receiver side satisfies the user requirement $\gamma$. The second constraint (\ref{eq:pha}) restricts $\Phi_n$ to be a discrete value from the set $\mathcal{F}$. Problem (\ref{eq:minp}) is a non-convex optimization problem, and so the work in \cite{Wu_BeamOpt} tackles the problem using an alternating optimization technique, where all $N$ phase shifts are optimized alternatively by tuning one phase shift at a time, while fixing the others. The authors in \cite{Wu_BeamOpt} analytically proved that IRS with discrete phase shifts can achieve the same asymptotic squared power gain of the continuous phase shifts.  It further emphasizes that utilizing discrete phase shifts accomplishes a considerable power saving.

 In \cite{SWIPT}, IRS are used to address the problem of simultaneous wireless information and power transfer (SWIPT). In a far-field, power transfer has low efficiency which limits the rate-energy trade-off performance of SWIPT systems. Furthermore, in SWIPT, energy harvesting receiver (EHR) demands high received power that is much higher than the information decoding receiver (IDR) power, which imposes practical efficiency issues. To overcome these challenges, the work in \cite{SWIPT} proposes a novel SWIPT system aided by IRS technology. This approach leverages the high beamforming gains achieved by the IRS to enhance the wireless power transfer efficiency and rate-energy trade-off performance of the SWIPT systems.

The system model  used in \cite{SWIPT} is shown in Fig. \ref{fig: APSWIPT}, which consists of a MISO IRS-aided SWIPT system from the AP to many receivers, i.e., IDRs and EHRs. The purpose of utilizing IRS is to enhance the efficiency of EHRs that are placed in the coverage area of the IRS. The IRS-assisted network has $N$ reflecting elements to support SWIPT from $M$ number of APs (with multiple antennas) to two types of receivers (each with a single-antenna), which are IDRs and EHRs, expressed by $\mathcal{K_I}$ and $\mathcal{K_E}$, respectively. Moreover, \cite{SWIPT} considers a quasi-static flat fading channel model to characterize the optimal rate-energy performance.
Note that the IRS can create an interference-free zone via passive beamforming and active beamforming at the IRS and at the AP, respectively. For simplicity, the work in \cite{SWIPT} assumes that the interference between the AP signals cannot be canceled by the IDRs. To address the SWIPT system limitations, the authors in \cite{SWIPT} aim at maximizing the EHRs' received weighted sum-power, while achieving a certain signal-to-interference-plus-noise ratio (SINR) threshold at IDRs. This is achieved by optimizing the transmit beamforming vectors and reflect phase shifts at the AP and IRS, respectively. The SINR of the $i$-th IDR is given as follows (by taking into consideration that IDRs can not cancel the interference caused by the energy signals):
\begin{equation}
    \text{SINR}_i = \frac{|\boldsymbol{h}_i^\dagger \boldsymbol{w}_i|^2}{\sum_{k\neq i, k\in \mathcal{K_I}} |\boldsymbol{h}_i^\dagger \boldsymbol{w}_k|^2 + \sum_{j\in \mathcal{K_E}} |\boldsymbol{h}_i^\dagger \boldsymbol{v}_j|^2 + \sigma_i^2}
\end{equation}

\noindent where $\boldsymbol{h}_i^\dagger = \boldsymbol{h}_{r,i}^\dagger \boldsymbol{\Theta} \boldsymbol{G} + \boldsymbol{h}_{d,i}^\dagger$ given that $\boldsymbol{h}_{r,i}^\dagger$, $\boldsymbol{h}_{d,i}^\dagger$, and $\boldsymbol{G}$ represent the channel links between the IRS and $i$-th IDR, AP and $i$-th IDR, and AP-IRS, respectively. Note that $\boldsymbol{w}_i$ and $\boldsymbol{v}_j$ are the precoding vectors for IDR and EHR, respectively, where $k,i \in \mathcal{K_I}$ and $j \in \mathcal{K_E}$. Here, let $\boldsymbol{S}$ be a positive semi-definite matrix that accounts for the energy weights of EHRs. The maximization problem can then be expressed as
\begin{maxi!}
		{\boldsymbol{w}_i,\boldsymbol{v}_j,\boldsymbol{\Phi}}    {\sum_{i \in \mathcal{K_I}} \boldsymbol{w}_i^\dagger \boldsymbol{S} \boldsymbol{w}_i + \sum_{j \in \mathcal{K_E}} \boldsymbol{v}_j^\dagger \boldsymbol{S} \boldsymbol{v}_j }
		{\label{eq:swpt}}{}
		\addConstraint{\text{SINR}_i \geq \gamma_i,~ \forall~ i \in \mathcal{K_I} }\label{eq::SINR}
		\addConstraint{\sum_{i \in \mathcal{K_I}} \|\boldsymbol{w}_i\|^2 + \sum_{j \in \mathcal{K_E}} \|\boldsymbol{v}_j\|^2 \leq P_{max}}\label{eq:po}
		\addConstraint{0 \leq \Phi_n \leq 2\pi, \forall~ n \in \mathcal{N}}\label{eq:ph}
	\end{maxi!}
where the received weighted sum-power by EHRs is given as
\begin{equation}
\sum_{j \in \mathcal{K_E}} \alpha_j E_j = \sum_{i \in \mathcal{K_I}} \boldsymbol{w}_i^\dagger \boldsymbol{S} \boldsymbol{w}_i + \sum_{j \in \mathcal{K_E}} \boldsymbol{v}_j^\dagger \boldsymbol{S} \boldsymbol{v}_j,
\end{equation}
\noindent and where constraint (\ref{eq::SINR}) assures that the SINR at different IDRs exceeds a certain threshold and (\ref{eq:po}, \ref{eq:ph}) express the power budget and phase shift constraints, respectively. In (\ref{eq:ph}),  $\mathcal{N}$ denotes the set of reflecting elements with cardinality |$\mathcal{N}$| = $N$. The problem (\ref{eq:swpt}) is a non-convex problem because of coupling the transmit beamforming vectors and IRS phase shifts in the objective function and in the SINR constraint. The optimization problem (\ref{eq:swpt}) is reformulated as an alternating optimization problem and then solved with semi-definite relaxation (SDR) by dropping the rank-one constraint, which is related to the transmit pre-coders. Finally, the transmit pre-coders are recovered through eigenvalue decomposition over the attained rank-one. The results in \cite{SWIPT} show that the SWIPT system with IRS can radically increase rate-energy performance.

In \cite{Beam_Wu}, the work proposes a new approach for point-to-point MISO wireless networks using passive IRS. IRS are used to assess the information transmitted from the AP to the user as shown in Fig. \ref{fig: APbasic}. Hence, the user jointly gets both the signals that are transmitted from the AP and the one reflected by the IRS. The main objective is to maximize the total received power at the receiver by optimizing the transmit beamforming and the phase shifts at the AP and at the IRS, respectively. The non-convex optimization problem is formulated as:
\begin{maxi!}
{ \boldsymbol{w}, \boldsymbol{ \Phi}}{ |(\boldsymbol{h}^\dagger_r{\boldsymbol{\Theta}{\boldsymbol{G}} +  \boldsymbol{h}^\dagger_d})\boldsymbol{w}|^2 }
{\label{one}}{}
{}{}
\addConstraint{||\boldsymbol{w}||^2 \leq P_{max}}{}
\addConstraint{0 \leq \Phi_n \leq2\pi,}~{}{\forall n=1, ..., N},
\end{maxi!}
\noindent The diagonal matrix $\boldsymbol{\Theta}=\text{diag}(\beta e^{j\Phi_1},...,\beta e^{j\Phi_n},...,\beta e^{j\Phi_N})$ accounts for both the amplitude reflection coefficient $\beta \in [0,1]$.
The problem (\ref{one}) is non-convex because of the coupled expression involving transmit beamforming {$\boldsymbol{w}$} and phase shifts {$\boldsymbol{\Phi} $} in the objective function. Therefore,  problem (\ref{one}) is reformulated using a centralized algorithm based on SDR to relax the rank-one constraint. The resultant problem is given as:
\begin{maxi}
{ \boldsymbol{V}}{ \text{tr} (\boldsymbol{RV})}
{\label{five}}{}
{}{}
\addConstraint{\boldsymbol{V}_{n,n} = 1, {{\forall} n=1,...,N+1}} {}
\addConstraint{\boldsymbol{V} \succeq 0}{} ,
\end{maxi}
where
\begin{equation}
\boldsymbol{R} =
\begin{bmatrix}
 \boldsymbol{\Psi}\boldsymbol{\Psi}^\dagger   &  \boldsymbol{\Psi} \boldsymbol{h}_d \\
\boldsymbol{h}_d^\dagger \boldsymbol{\Psi}^\dagger             & 0
\end{bmatrix}.
\end{equation}
$\boldsymbol{\Psi}=\text{diag}(\boldsymbol{h}_r^\dagger){\boldsymbol{G}}$ indicates the diagonal matrix of IRS-user link.
The authors in \cite{Beam_Wu} obtained the eigenvalue decomposition of $\boldsymbol{V}={\boldsymbol{U}\boldsymbol{\Sigma} \boldsymbol{U}^\dagger}$, where $\boldsymbol{U}=[e_1, ...e_{N+1}]$ is a unitary matrix, and $\boldsymbol{\Sigma} = diag(\lambda_1, ...,\lambda_{N+1})$ is a diagonal matrix, both with a size of $(N+1)\times (N+1)$.  The resultant problem in (\ref{five}) is a regular convex semi-definite program (SDP), and therefore, can be solved using CVX optimization solver \cite{boyd_book}.
Furthermore, the work in \cite{Beam_Wu} introduces a distributed algorithm with low-complexity to solve  (\ref{five}), where the transmit beamforming and phase shifts are tuned by the AP and IRS alternatively.

For the transmit beamforming vector $\boldsymbol{w}$, the objective function in (\ref{one}) provides the resulting inequality:
\begin{equation}\label{eq: tinequailty}
\begin{split}
|(\boldsymbol{h}\_r^\dagger \boldsymbol{\Theta}\boldsymbol{G} + \boldsymbol{h}\_d^\dagger)\boldsymbol{w}|  & = |\boldsymbol{h}\_r^\dagger \boldsymbol{\Theta}\boldsymbol{G} \boldsymbol{w} +  \boldsymbol{h}\_d^\dagger \boldsymbol{w}|\\
&\leq |\boldsymbol{h}\_r^\dagger \boldsymbol{\Theta}\boldsymbol{G} \boldsymbol{w}| +  |\boldsymbol{h}\_d^\dagger \boldsymbol{w}|
\end{split}
\end{equation}

\noindent Therefore, the equality in \eqref{eq: tinequailty} holds if and only if $\arg(\boldsymbol{h}\_r^\dagger \boldsymbol{\Theta} \boldsymbol{G}\boldsymbol{w})=\arg(\boldsymbol{h}\_d^\dagger\boldsymbol{w})=\varphi_0$. Also, by utilizing the change of variables, such as $\boldsymbol{h}\_r^\dagger \boldsymbol{\Theta}\boldsymbol{G}\boldsymbol{w}=\boldsymbol{v}^\dagger \boldsymbol{a}$, where, $\boldsymbol{v}=[e^{j\Phi_1},...,e^{j\Phi_N}]^\dagger $, $\boldsymbol{a}=\text{diag}(\boldsymbol{h}\_r^\dagger) {\boldsymbol{G} \boldsymbol{w}}$, and neglecting the constant term $|\boldsymbol{h}\_d^\dagger \boldsymbol{w}|$, the problem (\ref{one}) is reduced to (\ref{six}) as follows:
\begin{maxi}
{ \boldsymbol{v}}{ \boldsymbol{|v}^\dagger \boldsymbol{a}|}
{\label{six}}{}
{}{}
\addConstraint{{v_n} = 1, {{\forall n}=1,...,N}} {}
\addConstraint{\arg(\boldsymbol{v}^\dagger \boldsymbol{a}) = \varphi_0}{}
\end{maxi}
The optimal solution of (\ref{six}) is provided by  $\boldsymbol{v^*}=e^{j(\varphi_0-\arg(\boldsymbol{a}))}=e^{j(\varphi_0-\arg(\text{diag}(\boldsymbol{h}\_r^\dagger)\boldsymbol{G}\boldsymbol{w}))}$ as in  \cite{Beam_Wu}. Therefore, the identical $n$-th phase shift is can be written as
\begin{equation*}
\Phi_n^*= \varphi_0 - \arg(h^\dagger_{n,r} \boldsymbol{g}^\dagger_n\boldsymbol{w})
=\varphi_0 - \arg(h^\dagger_{n,r})-\arg(\boldsymbol{g}^\dagger_n\boldsymbol{w}).
\end{equation*}
\noindent Note that $h^\dagger_{n,r}$ is the $n$-th element of $\boldsymbol{h}^\dagger_r$ (and  $\boldsymbol{g}^\dagger_n$ represents the $n$-th row vector of $\boldsymbol{G}$). $\boldsymbol{g}^\dagger_n \boldsymbol{w}$ includes the transmit beamforming and the channel link between the AP and the IRS. Furthermore, the phase of $\boldsymbol{h}^\dagger_d \boldsymbol{w}$ is fixed as a constant for all iterations to allow a distributed implementation. Note that summation of the phase rotation and the beamforming vector is valid without adjusting the beamforming gain. Then the transmit beamforming can be written as
\begin{equation}
    \boldsymbol{w^{\ast}} = \sqrt{P_{max}} \dfrac{(\boldsymbol{h}^\dagger_{r}\boldsymbol{\Theta} \boldsymbol{G} + \boldsymbol{h}^\dagger_d)^\dagger}{||\boldsymbol{h}^\dagger_{r}\boldsymbol{\Theta} \boldsymbol{G} + \boldsymbol{h}^\dagger_d||}e^{j\alpha}.
\end{equation}
\noindent The AP adaptively chooses $\alpha$ in all the iterations such that $\boldsymbol{h}^\dagger_d \boldsymbol{w}^*$ is a real number. In conclusion, the solution of problem (\ref{six}) can be achieved by applying the appropriate $n$-th phase shift at the IRS. The distributed algorithm does not require a feedback channel between the AP-IRS as compared to the centralized algorithm. Also, it does not require utilizing SDP solution since closed-form solutions exist.

\subsection{Sum-rate Maximization}
Having discussed both EE and power consumption from an optimization perspective, we hereby review several optimization formulations of maximizing the data rate gains to fully exploit the IRS technology. In \cite{deeplearning},  all the IRS' elements are considered passive in the presence of a few active elements, which are controlled by  the IRS controller. The IRS discover the best way to interact with the incoming signal, provided the active elements, by using a deep learning-based solution. Furthermore, the main goal is to maximize the achievable rate, by designing the IRS reflection beamforming vector $\boldsymbol{w}$. The achievable rate is expressed as
\begin{equation}
R = \frac{1}{K} \sum_{k=1}^K \log_2 \left(1+\text{SNR}|(\boldsymbol{h}_{T,k}\odot \boldsymbol{h}_{R,k})^T\boldsymbol{w}|^2 \right),
\end{equation}
\noindent where $\odot$ denote the Hadamard product, $\text{SNR} = \frac{p_T}{K \sigma^2}$ represents the total transmit power over the noise, ${h_{T,k}}$ and  ${h_{R,k}}$ represent the downlink channels, and $K$ indicates the number of sub-carriers. The reflection beamforming vector in the IRS is created by using the RF phase shifter. The beamforming vector is chosen from a predefined codebook $\boldsymbol{Q}$ where the goal is to find an optimal transmit beamforming $\boldsymbol{w}^*$ that yields to an optimal rate as follows

\begin{equation}
    R^\ast = \underset{\boldsymbol{w} \in \boldsymbol{Q}}{\operatorname{max}}  \frac{1}{K}\sum_{k=1}^K \log_2 \left( 1 + \text{SNR } | (\boldsymbol{h}_{T, k} \odot  \boldsymbol{h}_{R, k})^T \boldsymbol{w}|^2   \right).
\end{equation}
A comprehensive search is required to look for the optimal beamforming reflected vector $\boldsymbol{w}^*$, which does not have a  closed-form solution because of the quantized codebook constraint and the time-domain exertion of the beamforming vector \cite{deeplearning}. This extensive search increases the complexity of hardware implementation and power consumption significantly. For that reason, the objective is to design IRS-aided systems based on a deep learning solution to find the optimal achievable rate while satisfying a low-training overhead and low hardware complexity. To this end, the work in \cite{deeplearning} proposes a deep learning-based solution to predict the optimal reflection matrix while satisfying a low training overhead. The authors have also proposed two other methods for the IRS system design with unknown channel knowledge; one is based on compressive sensing \cite{deep_compressed}, and the other is based on deep reinforcement learning \cite{deep_reinforcement}. All three methods leverage the proposed IRS architecture with few sparse active channel sensors.

Recently, the work in \cite{MISO} considers the system model illustrated in Fig. \ref{fig: APbasic} and proposes maximizing spectral efficiency of the IRS-based system using two different algorithms, including manifold optimization and fixed-point iteration methods. The proposed algorithms achieve higher spectral efficiency with lower computational complexity. The maximization problem is formulated as

\begin{maxi}
{ \boldsymbol{\Theta},\boldsymbol{w}}{ |(\boldsymbol{h}^\dagger_r{\boldsymbol{\Theta}}\boldsymbol{G} +  \boldsymbol{h}^\dagger_d)\boldsymbol{w}|^2 }
{\label{first}}{}
{}{}
\addConstraint{\boldsymbol{\Theta} = \text{diag}(e^{j\Phi_1},e^{j\Phi_2}, ..., e^{j\Phi_M)}}{}
\addConstraint{||\boldsymbol{w}||^2 \leq P_{max}}{},
\end{maxi}
Note that the above optimization problem (\ref{first}) is non-convex due to the presence of phase shifts, and can be re-written as
\begin{maxi}
{\boldsymbol{v} }{\boldsymbol{v}^\dagger \boldsymbol{R}\boldsymbol{v}}
{\label{max}}{}
\addConstraint{|\upsilon_i| = 1, i\in \{1, 2, ..., M+1 \}}{},
\end{maxi}
where $\boldsymbol{v}=[\boldsymbol{x}^T,t]^T$, $\boldsymbol{x}=[e^{j\Phi_1},..., e^{j\Phi_M}]^\dagger$, $t \in \mathbb{R}$, and
\begin{equation}
\boldsymbol{R} =
\begin{bmatrix}
\text{diag} (\boldsymbol{h}^\dagger_r) \boldsymbol{GG}^\dagger \text{diag}(\boldsymbol{h}_r) &  \text{diag}(\boldsymbol{h}^\dagger_r) \boldsymbol{G}\boldsymbol{h}_d\\
\boldsymbol{h}^\dagger_d \boldsymbol{G}^\dagger \text{diag}(\boldsymbol{h}_r)              & 0
\end{bmatrix}.
\end{equation}
Note that (\ref{max}) is a quadratically constrained quadratic program (QCQP) where the  objective function is concave and can be solved with an SDR method by discarding the rank-one constraint. This method provides an estimated solution, i.e., it does not guarantee an optimal solution. As mentioned before, the fixed point iteration method and manifold optimization can be utilized to find a locally optimal solution for problem (\ref{first}).

Several other works on IRS assume fully reflective ideal phase-shift models, where they assume a unified amplitude at any phase shift. This implementation is not practical, due to hardware limitations. In contrast, \cite{Intelligent_RIS} introduced a feasible phase shift model with a reflection coefficient which apprehends the phase-dependent amplitude for a MISO wireless system. The IRS controller is utilized to communicate with the AP to control the IRS reflections, where the IRS reflecting elements are programmable by the controller. Recall that $ \boldsymbol{\Psi} =\text{diag}(\boldsymbol{h}^\dagger_r)\boldsymbol{G}$.
To design an IRS-assisted system, it is crucial to identify the relationship between the reflection amplitude and phase shift. Let the incident signal be denoted as $v_n=\beta_n(\Phi_n) e^{j\Phi}$, where $\Phi_n$, and $\beta_{n}$ denote the phase shift and its amplitude, respectively. Based on these parameters, Fig. \ref{fig: APbasic} depicts the system model that is considered in \cite{Intelligent_RIS} to formulate the optimization problem that maximizes the achievable rate by optimizing both the transmit beamforming vector $\boldsymbol{w}$ at the AP and the reflect beamforming vector $\boldsymbol{v}$ at the IRS as follows
\begin{maxi!}
{ \boldsymbol{w},\boldsymbol{v},\{\Phi_n\}}{ |(\boldsymbol{v}^\dagger{\boldsymbol{\Psi}} +  \boldsymbol{h}^\dagger_d)\boldsymbol{w}|^2 } {\label{eq:IntRIS1}}
{}
\addConstraint{||\boldsymbol{w}||^2 \leq P_{max}}\label{eq:IntRIS2}
\addConstraint{v_n = \beta_n(\Phi_n)e^{j\Phi_n},~ \forall n = 1,...,N }\label{eq:IntRIS3}
\addConstraint{-\pi \leq \Phi_n \leq \pi,~ \forall n = 1,...,N, }\label{eq:IntRIS4}
\end{maxi!}

\noindent where (\ref{eq:IntRIS2}) denotes the power constraint at the AP. Constraint (\ref{eq:IntRIS3}) denotes the reflection amplitude as a function of the phase shift, whereas  (\ref{eq:IntRIS4}) accounts for the phase shift to be between -$\pi$ and $\pi$.  According to \cite{Intelligent_RIS}, the optimal transmit beamforming optimal transmit beamforming for \eqref{eq:IntRIS1} is found using the maximum-ratio transmission, where $\boldsymbol{w}^*=\sqrt{P_{max}}\frac{{((\boldsymbol{v}^\dagger{\boldsymbol{\Psi}} +  \boldsymbol{h}^\dagger_d)^\dagger)}}{||(\boldsymbol{v}^\dagger \boldsymbol{\Psi} +  \boldsymbol{h}^\dagger_d) ||}$. By accounting for $\boldsymbol{w}^*$, the problem in \eqref{eq:IntRIS1} is reformulated as
\begin{maxi!}
{ \boldsymbol{v},\Phi_n}{ ||(\boldsymbol{v}^\dagger{\boldsymbol{\Psi}} +  \boldsymbol{h}^\dagger_d)||^2 } {\label{eq:IntRIS12}}
{}
\addConstraint{v_n = \beta_n(\Phi_n)e^{j\Phi_n}, \forall n = 1,...,N}\label{eq:IntRIS32}
\addConstraint{-\pi \leq \Phi_n \leq \pi, \forall n = 1,...,N, }\label{eq:IntRIS42}
\end{maxi!}

Although problem \eqref{eq:IntRIS1}  is simplified in \eqref{eq:IntRIS12}, it remains non-convex and is complicated to be solved using classical techniques. Hence, the work in \cite{Intelligent_RIS}  utilizes an alternating optimization approach to tackle the problem, reaching a sub-optimal solution. In \cite{Weighted}, a weighted sum-rate (WSR) is maximized instead, by jointly determining the active beamforming at the BS and the passive beamforming at the IRS. A downlink multiuser MISO communication system is considered in \cite{Weighted} as shown in Fig.~\ref{fig: basic}, containing $M$ antennas at BS, IRS with $N$ reflecting elements, and $K$ end-users \cite{Weighted}.
The IRS are utilized to assist the BS in reducing the fading and shadowing effects, where the instantaneous SINR for decoding the intended signal at user $k$ is denoted as
\begin{equation}
\gamma_k= \frac{|(\boldsymbol{h}^\dagger_{d,k} + \boldsymbol{h}^\dagger_{r,k} \boldsymbol{\Theta}^\dagger \boldsymbol{H}_1)\boldsymbol{w}_k|^2}{\sum_{i=1, i\neq k}^K |(\boldsymbol{h}^\dagger_{d,k}} + \boldsymbol{h}^\dagger_{r,k}  \boldsymbol{\Theta}^\dagger \boldsymbol{H}_1)\boldsymbol{w}_i|^2 + \sigma^2.
\end{equation}
Here the $k$-th user treats signals that are transmitted from other users as interference. Hence, the optimization problem is formulated as
\begin{maxi!}
{\boldsymbol{W},\boldsymbol{\Theta}}{f_1(\boldsymbol{W},\boldsymbol{\Theta})= \sum_{k=1}^K w_k \log_2(1+\gamma_k)} {\label{eq:w1}}
{}
\addConstraint{\Phi_n \in \mathcal{F},~ \forall n = 1,...,N }\label{eq:w2}
\addConstraint{\sum_{k=1}^K ||\boldsymbol{w}_k||^2 \leq P_{max} }\label{eq:w3}.
\end{maxi!}
The goal in (\ref{eq:w1}) is to maximize the WSR as a function of $\boldsymbol{W}$ and $\boldsymbol{\Theta}$, where it represents the transmit beamforming matrix at the BS and the reflection coefficient matrix at IRS, respectively. The term $w_k$ indicates the priority of the $k$-th user. In (\ref{eq:w2}), $\mathcal{F}_1,  \mathcal{F}_2,  \mathcal{F}_3$ denote an ideal reflection coefficient, continuous, and discrete phase shifters,  respectively, where $\mathcal{F} \in \{\mathcal{F}_1,  \mathcal{F}_2,  \mathcal{F}_3\}$. In addition, transmit power constraint (\ref{eq:w3}) denotes the maximum feasible threshold power $P_{max}$. The authors in \cite{Weighted} tackled this non-convex optimization problem via Lagrangian dual transform with a low computational complexity algorithm, by alternatively optimizing the active and passive beamforming so as to attain a sub-optimal solution. The results suggest that beamforming optimization surpasses conventional models under given phase-shift models.

\subsection{Secrecy-rate Maximization} 
Besides improving the achievable rate and EE of wireless communication systems,  IRS can also achieve physical layer security. In fact, the IRS  improve the secrecy data rate, i.e., when the data rate at an eavesdropper decreases and the data rate at a legitimate receiver increases \cite{zhao2019survey, Dong2020}. Additionally, IRS focused transmissions from the smart meta-surfaces would boost communication network security. Therefore, given that signals might be subject to interception after transmission, optimizing secrecy rates becomes quite important. In \cite{Secrecy, qiao2020secure}, the secrecy rate of the system is maximized by optimizing the source transmit power and the IRS’ phase shift matrix. Since the formulated problem in \eqref{eq:SecRate} is not convex, the work in \cite{Secrecy} proposes an alternating algorithm to acquire a tractable solution. First, a closed-form solution for the source transmit power is obtained and then a bisection search based semi-closed form solution is developed via tight bounding to optimize the phase shift matrix.

\begin{figure}
    \centering
    \includegraphics[width =0.48\textwidth]{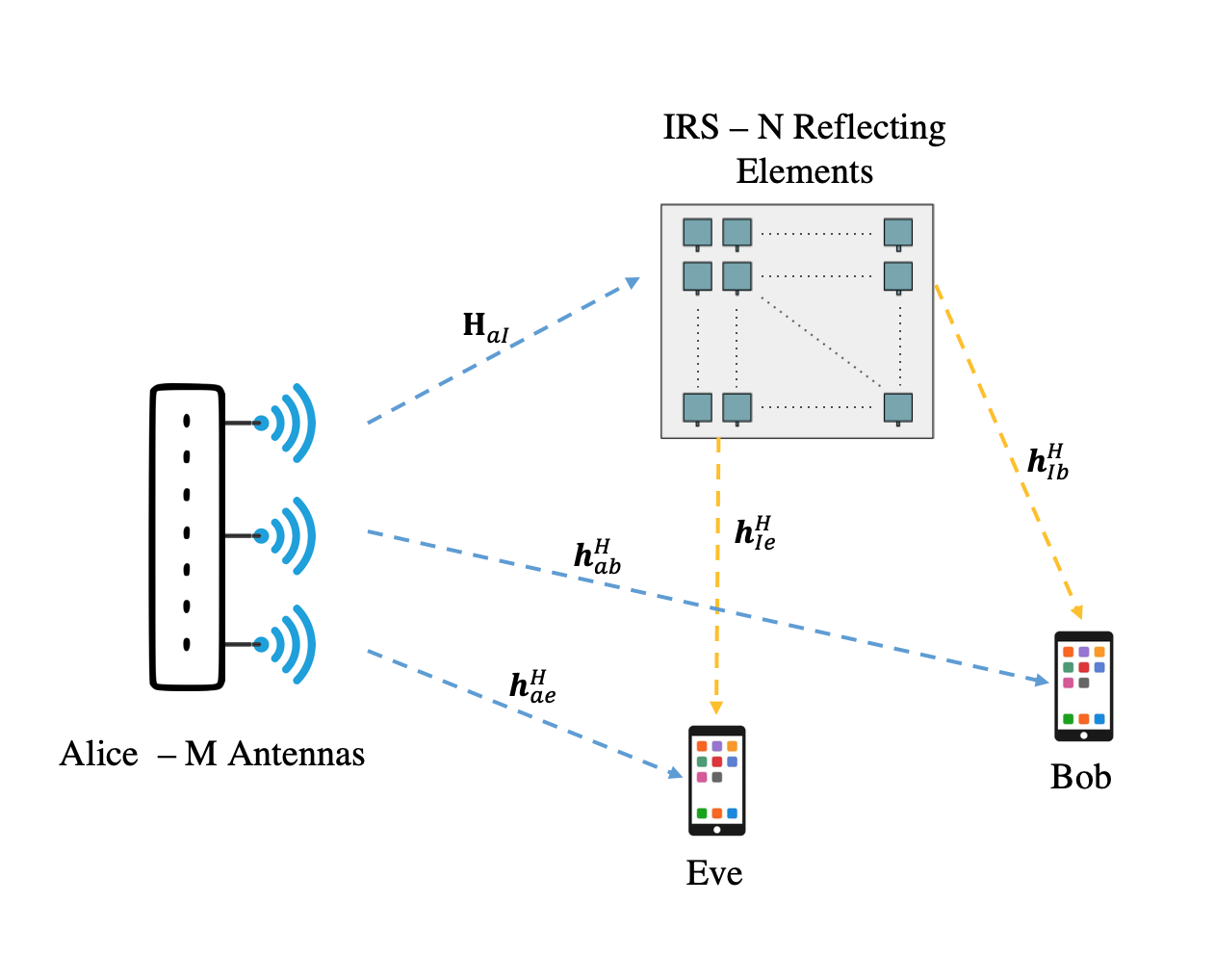}
    \caption{An IRS-aided wireless communication system subject to interception.\label{fig: SR}}
    \end{figure}

The system model in Fig.~\ref{fig: SR} consists of a source (Alice), IRS, receiver (Bob), and an eavesdropper (Eve). $M$ indicates the number of antennas at Alice's side, and  $N$ represents the number of reflecting elements at the IRS.
Also, reflected signals from the IRS are neglected due to their small power. Furthermore, to maximize the power matrix of the reflected signal, the work considers maximal reflection of the reflected signal matrix.
Due to the fact that the IRS reflect Alice's signal, the received signals at Bob's side are the combination of both the signal from Alice and the IRS. The signal received at Bob is denoted by
\begin{equation}
y_b = \boldsymbol{h}^\dagger_{Ib} \boldsymbol{\Theta} \boldsymbol{H}_{aI} \boldsymbol{x} + \boldsymbol{h}^\dagger_{ab}\boldsymbol{x}+ n_b,
\end{equation}
where $\boldsymbol{h}^\dagger_{Ib}$ is the  channel between Bob and the IRS. $\Phi_i$ denotes the phase shift obtained by the $i$-th reflecting element of the IRS, $\boldsymbol{H}_{aI}$ is the channel between Alice and the IRS, and $\boldsymbol{x}$ represents the transmitted signal by Alice. The co-variance matrix  $\boldsymbol{W}= \mathbb{E} \{\boldsymbol{xx}^H\}$ of $\boldsymbol{x}$ satisfies $\text{tr}(\boldsymbol{W}) \leq P_{max}$, where $\boldsymbol{h}^\dagger_{ab}$ is the channel between Alice and Bob, and $n_b$ is the AWGN with variance $\sigma^2_{n, b}$ at Bob's side. Similarly, the signal received by Eve is given as
 \begin{equation}
 y_e = \boldsymbol{h}^\dagger_{Ie} \boldsymbol{\Theta} \boldsymbol{H}_{aI} \boldsymbol{x} + \boldsymbol{h}^\dagger_{ae}\boldsymbol{x}+ n_e,
 \end{equation}
where $\boldsymbol{h}^\dagger_{Ie}$, $\boldsymbol{h}^\dagger_{ae}$ are the channel between the IRS and Bob, and the channel between Alice and Eve, respectively.  The AWGN at Eve is denoted by $n_e$ with variance $\sigma^2_{n,e}$.
To improve the secrecy rate for the above system setup,  $\boldsymbol{W}$ and the phase shift matrix $\boldsymbol{\Theta}$ are jointly optimized. The achievable rate ${{R_s (\boldsymbol{W} , \boldsymbol{\Theta})}}$ needs to be maximized, i.e.,
\begin{maxi!}
		{\ \boldsymbol{W} \succeq 0 \,\ , \boldsymbol{\Theta}}    {{R_s (\boldsymbol{W} ,\boldsymbol{\Theta})}}
		{\label{eq:SecRate}}{}
		\addConstraint{\text{tr}(W) \leq P, |\theta_i| = 1 , i = 1, \cdots , N},
	\end{maxi!}
where $R_s$ is given by
\footnotesize{
\begin{align}
{{R_s(\boldsymbol{W} ,\boldsymbol{\Theta})}} &=  \nonumber \\ &\log_2 \left(1+ \frac{(\boldsymbol{h}^\dagger_{Ib} \boldsymbol{\Theta} \boldsymbol{H}_{aI}  + \boldsymbol{h}^\dagger_{ab}) \boldsymbol{W} (\boldsymbol{H}^\dagger_{aI} \boldsymbol{\Theta}^\dagger \boldsymbol{h}_{Ib} + \boldsymbol{h}_{ab})}{\sigma^2_{n,b}}\right)\nonumber \\
& -\log_2 \left(1+ \frac{(\boldsymbol{h}^\dagger_{Ie} \boldsymbol{\Theta} \boldsymbol{H}_{aI} + \boldsymbol{h}^\dagger_{ae}) \boldsymbol{W} (\boldsymbol{H}^\dagger_{aI} \boldsymbol{\Theta}^\dagger \boldsymbol{h}_{Ie} + \boldsymbol{h}_{ae})}{\sigma^2_{n,e}} \right).
\end{align}
}
\normalsize
\noindent Note that the $i$-th diagonal element of $\boldsymbol{\Theta}$ is denoted by ${\Phi_i}$. Due to the non-convexity caused by the unit modulus constraints, the optimization problem \eqref{eq:SecRate} is hard to solve. Hence, to solve the secrecy rate maximization problem, an alternating algorithm approach is used \cite{Secrecy}. The algorithm optimizes $\boldsymbol{W}$ with fixed $\boldsymbol{\Theta}$, and alternatively optimizes $\boldsymbol{\Theta}$ for a given $\boldsymbol{W}$.  Optimization over $\boldsymbol{W}$ with a fixed $\boldsymbol{\Theta}$ yields the following problem
\begin{maxi!}
		{\boldsymbol{W}\succeq 0}    {{\frac{\boldsymbol{h}^\dagger_b \boldsymbol{W} \boldsymbol{h}_b + \sigma^2_{n,b}}{\boldsymbol{h}^\dagger_e \boldsymbol{W} \boldsymbol{h}_e + \sigma^2_{n,e}}}}
		{\label{eq:SecRateW}}{}
		\addConstraint{\text{tr}(\boldsymbol{W}) \leq P_{max}},
	\end{maxi!}
\noindent where $\boldsymbol{h}_b$ = $\boldsymbol{H}^\dagger_{aI} \boldsymbol{\Theta}^\dagger \boldsymbol{h}_{ab}$ and $\boldsymbol{h}_e = \boldsymbol{H}^\dagger_{aI} \boldsymbol{\Theta}^\dagger \boldsymbol{h}_{ae}$. Similarly, the problem that maximizes $\boldsymbol{\Theta}$ by fixing $\boldsymbol{W}$, is given by
\begin{maxi!}
		{\ \boldsymbol{\Theta}} {{\frac{|(\boldsymbol{h}^\dagger_{Ib} \boldsymbol{\Theta} \boldsymbol{H}_{aI}+ \boldsymbol{h}^\dagger_{ab})\boldsymbol{w}|^2 + \sigma^2_{n,b}} {|(\boldsymbol{h}^\dagger_{Ie} \boldsymbol{\Theta} \boldsymbol{H}_{aI}+ \boldsymbol{h}^\dagger_{ae})\boldsymbol{w}|^2 + \sigma^2_{n,e}}}}
		{\label{eq:SecRateTheta}}{}
		\addConstraint{|\Phi_i| = 1 , i = 1, \cdots , N}.
	\end{maxi!}
The optimization problem \eqref{eq:SecRateTheta} is solved by using fractional programming. The work in \cite{Secrecy} also developed a secrecy maximization algorithm for multi-antenna communication systems where Eve has $M \geq 1$ antennas.
\begin{table*}[ht]
\centering
\caption{Comparison of optimization frameworks for LIS.}\label{table: papcomp}
 \begin{tabular}{| >{\centering\arraybackslash}m{0.5in} | >{\centering\arraybackslash}m{6in} |}
 \hline
\textbf{Ref.} & \textbf{Assumptions \& Limitations}\\\hline\hline
\cite{energEff} &  Ignored the BS-user link due to unfavorable propagation conditions\\ \hline

\cite{Wu_BeamOpt} & Considered the reflected signals by the IRS for the first time only and ignored the rest. Also, the received signal from AP-user for asymptotically large $N$ is ignored\\ \hline

\cite{SWIPT} & Assumed perfect channel state information (CSI) at the AP. Also, the work assigns each receiver with only one dedicated energy beam\\ \hline

\cite{Beam_Wu} & Neglected the power signal that is displayed by the IRS more than two times. Moreover, a flat-fading channel model is assumed, and perfect CSI at the IRS \\ \hline

\cite{MISO} & Assumed a flat-fading channel model and perfect CSI at the AP and the IRS. Also, the work considers that all channels follows an independent Rayleigh fading\\
\hline
\cite{Intelligent_RIS} &  Assumed a quasi-static flat fading channel model, and considered reflected signals by the IRS for the first time only and ignored the rest\\ \hline
\cite{Weighted} & Assumed quasi-static flat-fading channels and known CSI  \\ \hline

\cite{Secrecy} & Considered maximal reflection without loss at the IRS\\ \hline

\cite{nadeem_asymptotic}& Assumed that the direct paths of the signal between the BS-users are blocked by obstacles and also perfect CSI is known at the BS. \\ \hline
  \end{tabular}
\end{table*}

\subsection{Coverage Optimization}
In the past, designing precoding and beamforming techniques for MIMO systems has received significant attention. This includes the minimization of transmit power  and maximization of the minimum SINR. Maximizing the minimum SINR has not been analyzed much by taking into consideration the reflect beamforming design of the IRS. Therefore, the work in \cite{nadeem_asymptotic} proposes optimizing the IRS phase matrix that maximizes the minimum SINR. A projected gradient ascent algorithm has been used to solve the optimization problem and determine the phases that maximize the minimum user SINR under optimal linear precoder (OLP). The multi-user MISO communication system is shown in Fig.~\ref{fig: basic}, which consists of a BS with $M$ antennas that communicate with $K$ single-antenna users, and IRS that has $N$ passive reflecting elements. The IRS are established on the surrounding building's wall that can adjust the phase shift of each reflecting component to realize the desired communication objective. In \cite{nadeem_asymptotic}, the work assumes that the BS-to-IRS channel satisfies the LoS conditions. Furthermore, it considers a complex scattering environment, spatial correlation between the IRS elements, and correlated Rayleigh channels between the IRS and the users due to the user's mobility. In \cite{nadeem_asymptotic},  the transmitted signal $\boldsymbol{x}$ is denoted by
\begin{equation}
\boldsymbol{x} =  \sum_{k=1}^K \sqrt{\frac{p_k}{K}} \boldsymbol{g}_k s_k,
\end{equation}
where $\boldsymbol{g}_k \in \mathbb{C}^{M \time 1}$ is the precoding vector, $p_k$ is the signal power, and $s_k$ is the data symbol for the $k$-th user, respectively. Moreover, the transmitted signal satisfies the constraint of the average transmission power per user. Hence, the downlink SINR at a single user $k$ is given by
\begin{equation}
\gamma_k =
\frac{\frac{p_k}{K}|\boldsymbol{h}^\dagger_k \boldsymbol{g}_k|^2}{\sum_{i\neq k}\frac{p_i}{K}|\boldsymbol{h}^\dagger_k \boldsymbol{g}_i|^2 + \sigma^2}
\end{equation}
here $\boldsymbol{h}_k$=$\sqrt{\beta_k}\boldsymbol{H}_1 \boldsymbol{\Theta}\boldsymbol{R}_{\text{IRS}_k}^{1/2}\boldsymbol{h}_{r,k}$ denotes the overall channel between the BS and user $k$, where $\boldsymbol{R}_{\text{IRS}_k}$ denotes the spatial correlation
matrix of the IRS with respect to user $k$. The max-min SINR problem is then defined as,
\begin{maxi!}
		{\ \boldsymbol{P,G}} {\min\limits_{k} \hspace{0.5cm} \gamma_k}
		{\label{eq:SINR}}{}
		\addConstraint{{\frac{1}{K} \boldsymbol{1}^T_K \boldsymbol{P} \leq P_{max}}}
		\addConstraint{\parallel\boldsymbol{g}_k\parallel = 1,~  \forall k.}\label{eq:users}		
	\end{maxi!}
	
 This optimization problem in \eqref{eq:SINR} can be solved using the OLP to maximize the minimum SINR \cite{sinr_g}. The aim is to design the phase values of the IRS' elements that appear in the diagonal of the reflect beamforming. To achieve this goal,  an approach with infinite resolution phase shifters where all channels are precisely known at the BS was assumed in \cite{sinr_g}, which employs the OLP to allocate the optimal powers.

To summarize, when the LoS channel between the BS and the IRS is of rank-one, a closed-form solution with minimum SINR under the OLP is formulated \cite{sinr_g}. As $K$ increases, serving more than one user becomes more challenging due to the SINR convergence. To solve this problem, the work \cite{sinr_g} assumed that the LoS channel has a high rank and used the tools from random matrix theory developing a deterministic approximation for the OLP parameters \cite{nadeem_asymptotic}. Lastly, the IRS phase matrix that maximizes the minimum SINR under the asymptotic OLP was designed using projected gradient ascent.

This section explores relevant optimization frameworks so as to exploit the full capabilities of IRS-based systems. As a summary, we also provide Table. \ref{table: papcomp} which compares the optimization frameworks for IRS and states the assumptions considered in the literature. The next section focuses on analyzing the performance of the IRS-based system from several perspectives including capacity, hardware  impairments, and data rate.

\section{Performance Analysis of  LIS Systems}\label{sec:perf}
LIS technology is expected to provide reliable wireless communication when a LoS link is established, as previous sections discuss. The significance of this technology makes it vital to study the performance analysis of LIS-based systems. To this end, this section reviews the performance analysis of LIS systems from different aspects, such as asymptotic analysis of uplink and downlink data rate, outage probability, and spectral efficiency. Because of their relatively large dimensions, it is often challenging to get closed-form solutions to  mathematically describe LIS systems. This section, therefore, covers different methods of estimating and approximating LIS systems performance.

\subsection{Capacity Analysis of LIS Systems }
Due to LIS capabilities in empowering robust and high-speed 6G communication networks, it is evidently essential to study the systems capacity, including hardware impairment effect, uplink/downlink transmission rate, and the impact of phase shifts. 

\subsubsection{Data Transmission Capacity}
The authors in  \cite{Hu_datatx} study the achieved unit-volume normalized capacity, where an infinitely-sized LIS system and a fixed transmit power per unit volume $P_u$ are considered. As the wavelength $\lambda$ goes to zero,  the normalized capacity per unit-volume $\hat{C}$ approaches $\frac{P_u}{2N_0}$, $N_0$ is the power spectral density (PSD) of AWGN, and $P_u = \frac{P}{\lambda}$, where $P$ refers to the terminal transmit power.
The received signal in \cite{Hu_datatx}'s system is optimized and goes through a  sinc-function-like match filtering process. As for the analysis, the LIS system's spatial degree of freedom (DoF) is also studied in \cite{Hu_datatx} to be harvested (i.e., the number of independent signal dimensions $\rho$) \cite{DoF}, to align with Shannon's capacity. The spatial DoF is measured as a function of $\hat{C} $ for the high SNR slope, given as
 \begin{equation}\label{eq: DoF}
    \rho = \lim_{\frac{P_u}{2N_0} \to \infty} \dfrac{\hat{C}}{\log({\frac{P_u}{2N_0})}}.
 \end{equation}
For a one-dimensional user-equipment (UE) deployment, the normalized capacity is given by
 \begin{equation}\label{eq: C1}
    \hat{C} = \lim_{K \to \infty} \dfrac{C}{\Delta_x},
 \end{equation}
where $C,~K$, and $\Delta_x$ are the capacity,  number of UEs, and the spacing between the neighboring UEs on the $x$-dimension, respectively. Similarly, the normalized capacity for two-dimensional UE deployment is given by
 \begin{equation}\label{eq: C2}
    \hat{C} = \lim_{K \to \infty} \dfrac{C}{\Delta_s},
 \end{equation}
where $\Delta_s = \Delta_x \Delta_y $ is the spacing between the neighboring UEs.  $\Delta_x$ and $\Delta_y$ represent the spacing between the neighboring UEs on the $x$- and $y$-dimension respectively. Solving (\ref{eq: C1}) and (\ref{eq: C2}) boils to having $ \Delta_x = \frac{\lambda}{2}$ and $\Delta_s = \frac{\lambda}{\pi}$, which tells that for infinitely-sized LIS of one-dimension, $\hat{C}$ is maximized when $\frac{2}{\lambda}$ UEs are to be multiplexed per meter, whereas multiplexing $\frac{\pi}{\lambda^2}$ UEs spatially per meter$^2$ maximizes $\hat{C}$ for two-dimensional UE deployment.  Finally, \cite{Hu_datatx} leverages the sampling theory to show that a hexagonal lattice is an optimal sampling one, which minimizes the surface area of LIS when every deployed antenna can earn only a single spatial dimension. Also, increasing $K$ does not affect the capacity of the system, proving that it is a robust system with good potentials of data transmission interference suppression.
\subsubsection{Hardware Impairments Analysis}
LIS systems are believed to outperform the conventional communication systems; therefore, it is noteworthy to consider the analysis of LIS-systems in the presence of hardware impairments (HWI).  Almost all communication systems encounter HWI, and LIS systems are no exceptions. That is, LIS systems have large surface areas that would impact the HWI degradation over the system. The work in \cite{Capacity_hu} studies the capacity degradation of an LIS system in the presence of HWI, when serving one UE. The HWI is modeled as a Gaussian process, and its variance depends on $r$, the distance of received signal power from the center of the LIS, which is turn is modeled as
\begin{equation}
    f(r) = \alpha r^{2\beta},
\end{equation}
where $\alpha$ and $\beta$ are two positive constants, and where $\alpha = 0$ represents the case of no HWI. The capacity of an LIS system typically increases with a large surface area; however, an LIS system with HWI has the opposite performance. That is, increasing the surface area degrades the system capacity severely, as shown in the numerical results of \cite{Capacity_hu}. To overcome the significant impact of HWI on the LIS system, \cite{Capacity_hu} proposes splitting the LIS system surface into $K$ smaller units. This idea is yet to be discussed in section \ref{sec: position}.

\subsection{Data Rate Analysis}
One of the critical factors of analyzing communication systems is the achievable data rate. This section presents the literature on the data rate expressions derived for LIS systems.
 \begin{figure}
    \centering
    \includegraphics[width = \linewidth]{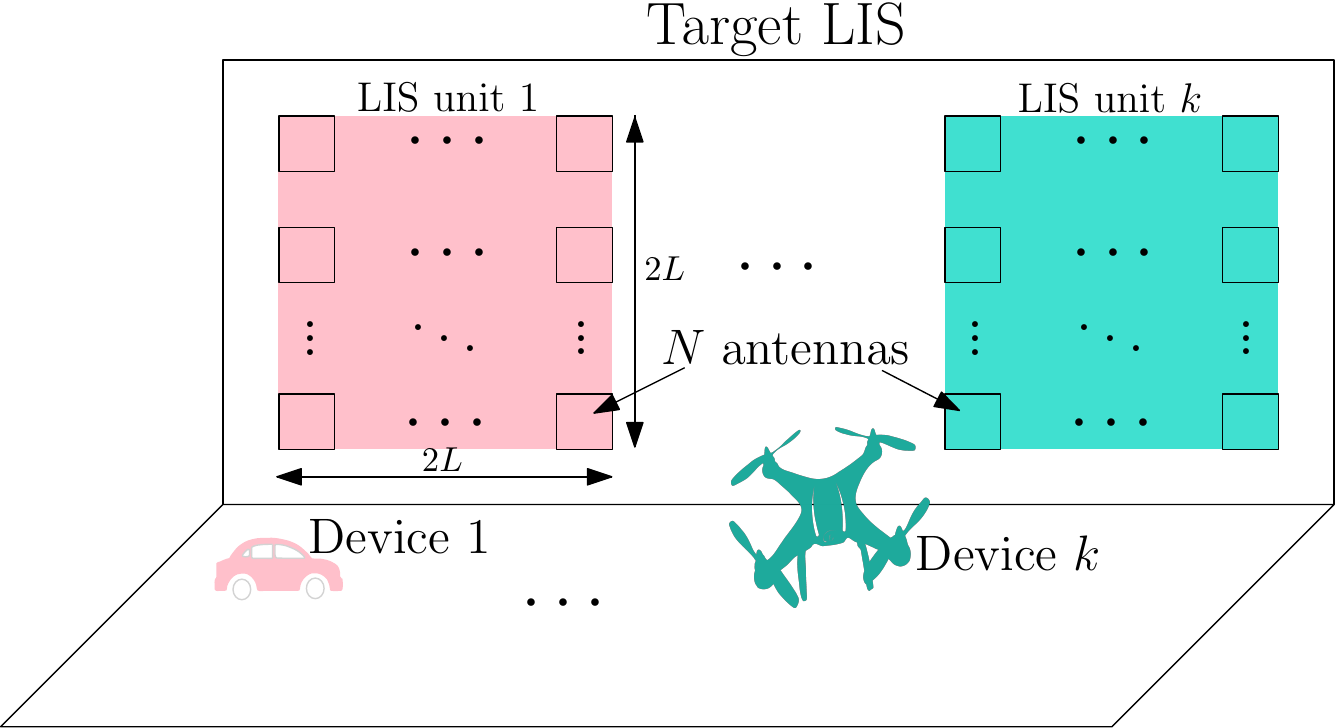}
    \caption{Single LIS system with $K$ units serving up to $K$ devices.\label{fig: single_LIS}}
\end{figure}
\subsubsection{Uplink Rate for single-LIS Systems}
The work in \cite{jung_datarate} studies not only the uplink data rate of LIS systems, but also shows the superiority of LIS systems' performance over mMIMO systems. Also, \cite{jung_datarate} takes into consideration the channel estimation error and channel hardening effects. In simple terms, \cite{jung_datarate} shows that the achieved capacity is correlated to the mutual information under the asymptotic analysis on the number of antennas and connected UEs.

The authors in \cite{jung_datarate} consider an LIS system shown in Fig. \ref{fig: single_LIS}. The system consists of a single large surface divided into a subarea of $2L\times2L$ denoted as units, each with $N$ antennas spaced by $\Delta L$  in a rectangular lattice. The LIS system serves up to $K$ devices, where each unit serves its corresponding device. To avoid performance degradation that may occur due to the overlapping between the LIS unit and the location of the device, \cite{jung_datarate} assumes deploying orthogonal resource management among devices. Therefore, all devices communicate to a non-overlapping unit.

Using the above system, \cite{jung_datarate} analyzes the asymptotic performance of the uplink data rate for boundless increasing $N$ and $K$, where \cite{jung_datarate} approximates the system performance by first defining the uplink data rate for one unit (i.e., unit $k$) as
 \begin{equation}
     R_k =  \log(1 + \gamma_k),
 \end{equation}
where $\gamma_k$ is the received SINR of unit $k$, and is given by
 \begin{equation}\label{eq: SINR_datarate}
     \gamma_k = \dfrac{\rho_k S_k (1 - \tau_k^2)}{I_k}.
 \end{equation}
 In (\ref{eq: SINR_datarate}), $\rho_k$ is the uplink transmit SNR of device $k$, $\tau_k \in [0, 1]$ represents the imperfection of the channel error estimation. The desired signal power is denoted as  $S_k = |\boldsymbol{h}_{kk}|^4$, $\boldsymbol{h}_{kk}$ is the channel between device $k$ and the unit $k$ of the LIS, and $I_k$ is the interference-plus-noise term for unit $k$.

 The authors in \cite{jung_datarate} approach the asymptotic performance analysis by first analyzing the moments of the random variable $I_k$, then, obtaining its asymptotic moment, and finally deriving the asymptotic moment of $R_k$ from $I_k$. The study \cite{jung_datarate} shows that $S_k$ converges to a constant that depends on the height of the device $k$ and length of the LIS unit when $N$ approaches infinity. $R_k$ mean and variance can be approximately derived based on the mean and variance of $\gamma_k$, denoted by $\mu_{\gamma_k}$ and $\sigma_{\gamma_k}^2$, respectively. Therefore, using Taylor expansion, the mean $\bar{\mu}_{R_k}$ and variance $\bar{\sigma}_{R_k}^2$ of $R_k$ can be defined exclusively by the random variable $I_k$, where the asymptotic mean and variance of $I_k$  depend on the devices' locations and the Non-Line-of-Sight (NLoS) interference correlation matrix:
 \begin{equation}
      \bar{\mu}_{R_k} = \log(1 + \bar{\mu}_{\gamma_k}) - \dfrac{\bar{\sigma}_{\gamma_k}^2}{2(\mu_{\gamma_k} + 1)^2},
 \end{equation}
 and
  \begin{equation}
      \bar{\sigma}_{R_k}^2 =   \dfrac{-\bar{\sigma}_{\gamma_k}^4}{4(\bar{ \mu}_{\gamma_k} + 1)^4} + \dfrac{\bar{\sigma}_{\gamma_k}^2}{(1 +\bar{ \mu}_{\gamma_k})^2},
 \end{equation}
 respectively, $\bar{\mu}_{\gamma_k}$ and $\bar{\sigma}_{\gamma_k}^2$ are the asymptotic mean and variance of $\gamma_k$. Hence, the evaluation of the LIS system performance can be obtained with no need to run extensive simulations.

 In large antenna-based systems, the system validity, latency, and diversity scheduling are controlled by the fluctuations of mutual information. For this reason, the study of the channel hardening effect is crucial for LIS-based systems. Hence, \cite{jung_datarate} studies the performance of mutual information variance as $N$ increases. Given the asymptotic value of $\gamma_k$ and $N = (\frac{2L}{\Delta L})^2$, the mean and variance of $\gamma_k$ can be expressed as the asymptotic mean $ \bar{\mu}_{I_k}/N^2$ and the asymptotic variance $ \bar{\sigma}_{I_k}^2/N^2$ of the random variable $I_k$. Using the scaling law, \cite{jung_datarate} shows that the asymptotic interference-plus-noise  $\bar{I_k}$ converges to a constant as $N$ increases; hence, the asymptotic data rate $\bar{R_k}$, its mean and variance converge to a constant value, proving that an LIS system indeed is influenced by the channel hardening effect. Based on that,  LIS lack scheduling diversity and have better reliability and latency because of the deterministic data rate. Finally, \cite{jung_datarate} compares the asymptotic performance of the LIS system with  mMIMO in terms of data rate's ergodic value and variance where the LIS system has a higher ergodic rate. As $N$ increases, however, the gap between the two starts shrinking, especially as the number of devices $K$ grows. Although the two systems have similar performance for large values of $N$, this solution is not practical for  mMIMO systems, as it requires a large physical area, unlike the LIS systems. In terms of data rate variance, a  mMIMO system shows a reduced channel hardening effect when the variance increases with $N$ and then converges to a constant. As for the LIS system, the variance goes to zero with an increase of $N$, which illustrates the channel hardening effect and eventually shows LIS systems performance superiority in terms of reliability and latency.
  \begin{figure}
     \centering
     \includegraphics[width =\linewidth]{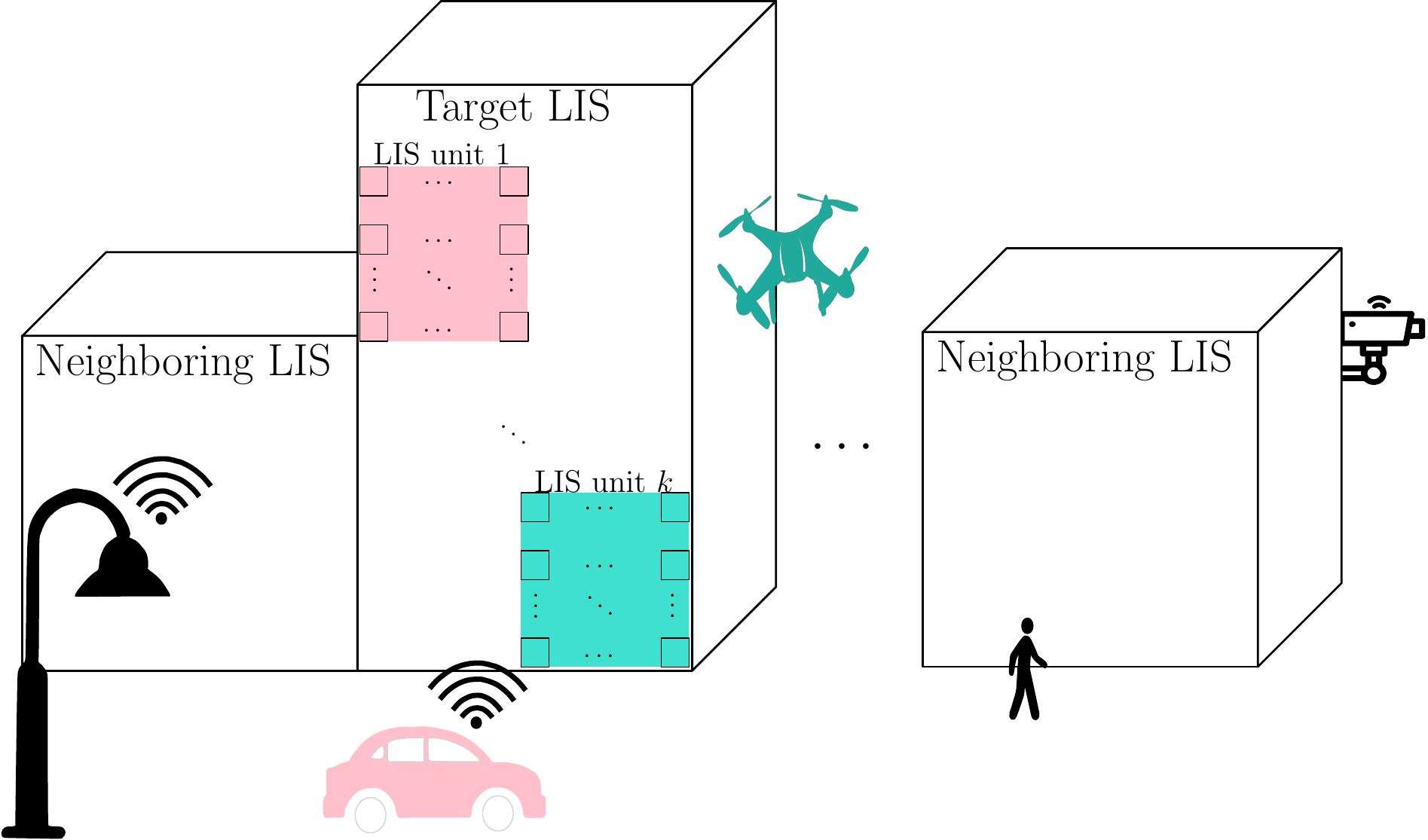}
     \caption{The system model considered in \cite{SSE}, illustrating a multi-LIS system sharing the same frequency band. \label{fig: multiLIS}}
     
 \end{figure}

\subsubsection{Uplink Rate for Multi-LIS Systems}
Unlike prior works that only study the performance of a single LIS system, \cite{SSE} studies the performance of a multi-LIS system (illustrated in Fig. \ref{fig: multiLIS}), by deriving an upper bound on the asymptotic system spectral efficiency (SSE) and by investigating the impact of pilot contamination. Based on the derivation of the upper bound, \cite{SSE} attempts to optimize the length of pilot training and the number of served devices per an LIS system. The significance of \cite{SSE}'s study lies in the fact that acquiring the CSI requires sending pilot signals; however, in multi-LIS systems, pilot contamination may occur due to inter-LIS interference.

\begin{figure}
     \centering
     \includegraphics[width = 0.8\linewidth]{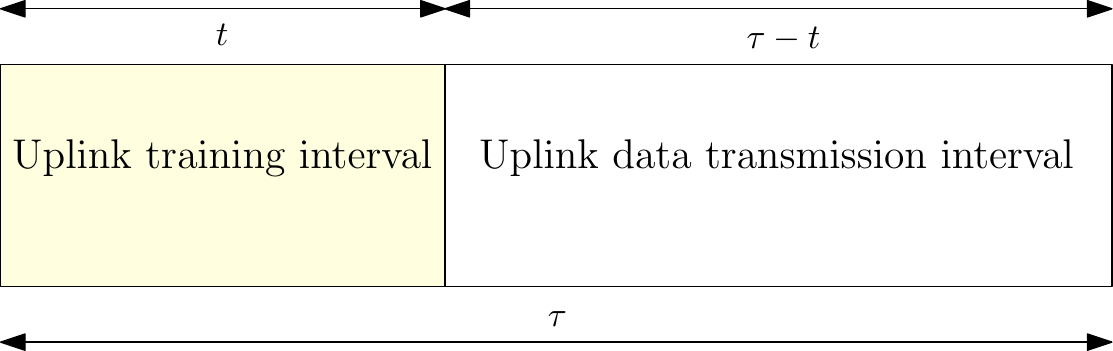}
     \caption{The uplink frame structure showing the interval period of a pilot signaling  and data transmission \cite{SSE}. \label{fig: uplink}}
 \end{figure}

The study in \cite{SSE} considers a multi-LIS system consisting of $Z$ number of LIS, each with the same characteristics as in \cite{jung_datarate}, where each surface of LIS has its signal processing module to receive signals, estimate CSI and detect uplink signal from its corresponding device.
The modeling of \cite{SSE}'s system ensures not having an overlapping LIS unit (i.e., no intra-LIS interference) by having an orthogonal multiple access resource management schemes for devices with similar locations.  Considering an LIS system with a matched filter (MF), the MF requires an accurate CSI to suppress the interference of a signal, where it gathers the CSI through pilot signaling from a device to its corresponding LIS. Pilot signaling of a device occurs during the coherent channel time $\tau$ within the uplink frame structure. In the uplink structure, pilot signaling occurs in a period of $t$, while data transmission takes a period of $\tau-t$, as shown in Fig.\ref{fig: uplink}. Given the orthogonal multiple access scheme deployed, an orthogonal pilot sequence of $t \geq K$ is allocated for each device for obtaining the required CSI at the LIS side. In the case of a multi-LIS system, the pilot symbols used by two adjacent LIS lose their orthogonality, causing pilot contamination.
For an uplink frame structure, the instantaneous SSE of the $z$-th LIS is given by
 \begin{equation}
     R_{z}^{SSE} = (1- \frac{t}{\tau}) \sum_{k=1}^K R_{zk} = (1- \frac{t}{\tau}) \sum_{k=1}^K \log(1 + \gamma_{zk}),
 \end{equation}
 where $R_{zk}$ and $\gamma_{zk}$ are the data rate and SINR of unit $k$, respectively. Next,  \cite{SSE} optimizes $t$, which maximizes the asymptotic SSE for the unboundedly growing $N$. As \cite{Hu_datatx} and \cite{jung_datarate} show, the signal power of unit $k$ of LIS $z$ (for $z \geq 1$) converges to a deterministic value, with the increase of $N$. Following the investigation of \cite{jung_datarate}, \cite{SSE} shows that unlike  mMIMO systems, as $N$ increases, a multi-LIS system has negligible inter/intra-LIS-interference through NLOS. Also, the imperfect CSI of a multi-LIS system does not affect its SSE; the SSE of a multi-LIS system, regardless of its CSI, achieves the same performance of a single-LIS with perfect CSI. Also, \cite{SSE} shows that the pilot contamination bounds a multi-LIS SSE performance due to the inter/intra-interference caused by LoS paths. Moreover, the authors in \cite{Zheng2020} analyze the performance of LIS using orthogonal multiple access (OMA) and non-orthogonal multiple access (NOMA). The study in \cite{Zheng2020}  reveals that NOMA may perform worse than OMA for the users nearby LIS.

 Unlike  mMIMO systems where $t$ is a factor affecting the received SINR, \cite{SSE} shows that when optimizing $t$ for LIS systems, increasing $t$ does not increase the SINR. Therefore, the optimal $t^{\ast}$ which maximizes the SINR is the minimum $t$ (i.e., $t = K$). Finally, \cite{SSE} asymptotically derives the optimal number of devices that maximizes SSE. Also, \cite{SSE} verifies the analytical derivations (approximation) with extensive simulations, where the study shows that the channel hardening effect of multi-LIS is closer to that of a single-LIS system, for a high number of $K$. The ergodic uplink rate of the single and multi-LIS system is shown, where the single-LIS system has the superior performance, and the gap between the two follow the analytical derivation that is a result of the generated pilot contamination and inter-LIS interference. For the SSE, the increase of inter-LIS interference causes $K$ to increase, which explains the increase in the gap between the two performances.

\subsubsection{Rate Impact on Phase Shifts }
As iteratively stated throughout this manuscript, controlling phase shifts in LIS systems is a crucial factor for fine-tuning the communication quality-of-service. According to \cite{limited},  the practical implementation of LIS relies on the limitations of phase shifts, which degrades the overall system performance. Therefore, \cite{limited}  studies the uplink assisted communication system performance and provides an approximated expression for the attainable data rate. Also, the work in \cite{limited}  shows the optimal number of phase shifts needed for a particular data rate threshold.
In an LIS assisted communication model that consists of a  planar array of $N$ electrically controlled elements, the number of phase shift patterns that can be generated by the LIS model is  $ 2^{u} $, where  $ u $  is the number of coding bits. The phase shifts have a uniform interval expressed by  $\Delta  \theta =\frac{ 2\pi }{2^{u}}$.  One can obtain the  phase shift value by multiplying the phase shift interval by an integer $s_{i, j}$ that satisfies $ 0 \leq s_{i, j} \leq 2^{u}-1$, i.e., the phase shift value is $ s_{i, j} \Delta  \theta$, where the subscripts $i, j$ refer to the element in the $i$-th row and $j$-th column, respectively. 
The number of phase shifts is limited in practice; thus, it is important to study the effect of phase shifts limitations on the reliability of the system. The phase shift error can be expressed as a function of the optimal phase shift $ \theta _{i, j}^{\ast}$ and the closest phase shift $\hat{ \theta }_{i, j}$, that is expressed as
\begin{equation}
 \delta _{i, j}= \theta _{i, j}^{\ast}-\hat{ \theta }_{i, j},
\end{equation}
where
\begin{equation}
 -\frac{2 \pi }{2^{u+1}} \leq  \delta _{i, j}<\frac{2 \pi }{2^{u+1}}.
\end{equation}
 To evaluate the data rate degradation,  $  \epsilon  $  must be defined, which is the ratio of the error caused by the limited phase shifts to the continuous phase shifts. Therefore, for the system performance to exceed a certain threshold, $ \epsilon  $  must be greater than  $ \epsilon _{0} $, i.e.,
\begin{equation}
\epsilon =\frac{\log_{2} \left( 1+ \mathbb{E} \left[ \hat{ \gamma } \right]  \right) }{\log_{2} \left( 1+ \mathbb{E} \left[  \gamma  \right]  \right) }  \geq  \epsilon _{0},
\end{equation}
where  $ \epsilon _{0} < 1 $,  $ \hat{ \gamma } $  is the SNR expectation, and  $ \gamma  $  is the received SNR. The attainable data rate expression can be found in  \cite{limited}, where the final expression is bounded from both sides. The upper bound is found when  the Rician factor $ \kappa  \rightarrow \infty $, while the lower bound is calculated when  $  \kappa  \rightarrow 0 $. The results in \cite{limited} show that the data rate increases with the increase of the LIS size. When the size of the LIS is sufficiently large, the SNR becomes proportional to the square of the number of LIS elements. 
Most importantly, \cite{limited} finds that the required number of bits to generate various phase-shifts depends on the size of the LIS in the Rician channel conditions. The numerical results indicate that three bits are needed for small-sized LIS, two bits for moderate size, and one bit for infinitely-sized LIS, which implies that two phase-shifts on average (i.e., $2^1=2$ phase-shifts) are enough for deploying extremely sizable LIS.

\section{Reliability Analysis of LIS}\label{sec:RelAnalysis}
While the above studies focus on the asymptotic analysis of LIS systems, it is equally important to reflect on the reliability of the LIS system from an error analysis perspective. This section, therefore, presents some of the works which study the error performance of  LIS systems.

\subsection{Rate Distribution and Outage Probability}
Since the coverage and interference levels of indoor networks depend on the location and properties of objects and obstacles, intelligent surfaces are competent enough to regulate the smart propagation environment. Henceforth, LIS provide a better quality of coverage, level of service, and improve  system performance. The improvement of coverage can be based on using frequency selective surfaces, and applying well-chosen machine-learning control algorithms \cite{inproceedingsSubrt}.

 In \cite{rel}, the authors attempt to characterize the coverage in terms of outage probability, which is a significant performance measure to estimate the reliability of the LIS systems. In \cite{rel}, the asymptotic analysis of the sum-rate is used to obtain the analytical expression of the outage probability. The study in \cite{rel} claims that the approximation provides a precise estimate of the probability of outage and reduces the necessity for extensive simulations and extraneous computational time. Moreover, the simulation results prove that despite the fluctuating SNR,  the probability of outage is unaffected when the number of antenna and devices is significantly large.
Using the same system model as in \cite{jung_datarate}  and \cite{SSE}, the authors in \cite{rel} formulate the sum-rate as follows:
\begin{equation}
\label{sumrate}
 R = \sum_{k=1}^{K}R_k.
\end{equation}
 Furthermore, \cite{rel} argues that because the individual rates ${R_1,R_2...R_K}$ are not identical in distribution, the distribution of the sum-rate cannot be defined. The derivation of the outage probability is, therefore, non-trivial. Instead of analyzing the individual rate, each rate can be written as a function of a random variable $I_k$, i.e.,
\begin{equation}
\label{sumrateI}
R_k = a_k + b_k I_k,
\end{equation}
where $a_k$ and $b_k$ are deterministic values that depend on the length of the LIS units. The distribution of $R$ can be found using the central limit theorem, such  that for large values of $N$ and $K$, the asymptotic distribution of $R$ can be estimated to follow a Gaussian distribution with mean and variance as:
\begin{equation}
\label{mean1}
\bar{\mu}_R= \sum_{k}\log{(1+ \dfrac{\rho_k\bar{p_k}(1-\tau_k^2)}{\bar{\mu_{I_k}}})}
\end{equation}
and
\begin{equation}
\label{var1}
\bar{\sigma}_R= \sum_{k} \dfrac{\bar{\sigma}^2_{I_k}\rho_k^2\bar{p}_k^2(1-\tau_k^2)^2}{\bar{\mu}_{I_k}^2(\bar{\mu}_{I_k}+\rho_k\bar{p}_k(1-\tau_k^2))^2}
\end{equation}
respectively. In (\ref{mean1}) and (\ref{var1}), $\bar{p_k}$ depends on the device location and $\rho_k$ denotes the transmit SNR. Finally, the closed-form expression of the probability of outage is given as
\begin{equation}
\label{prob1}
P_o= Pr[R < R_D]= 1- Q\left(\dfrac{R_D-\bar{\mu_R}}{\bar{\sigma_R}}\right),
\end{equation}
where $R_D$ is the sum-rate threshold, and $Q(\cdot)$ represents the $Q$-function.

\subsection{Probability of Error for Intelligent and Blind Transmission}
In \cite{Basar_Frontier}, the author provides a mathematical framework that studies the relationship between the reflecting elements, blind phases, and modulation errors in LIS systems. The work in \cite{Basar_Frontier} assumes having an LIS system with $N$ reconfigurable reflecting elements and studies the error performance for two scenarios. The first scenario considers an intelligent transmission at the LIS (i.e., the channel phases are known), and the second considers a blind transmission. By first deriving the SNR, one can get the symbol-error-rate (SER) for $M$-ary communication using phase-shift-keying (PSK). The author in \cite{Basar_Frontier} first compares the binary PSK with a pure AWGN signal and studies the effect of increasing $N$.

Using numerical evaluation, \cite{Basar_Frontier} shows that an LIS-based signal, which smartly adjusts the phases of the reflector (scenario-I), has low error-probability, even at low SNR values. Also, \cite{Basar_Frontier} shows that doubling $N$ improves the error performance with a $6$ dB gain. For the second scenario (scenario-II), where the channel phases are not known for the LIS, \cite{Basar_Frontier} shows that a gain of $N \times \text{SNR}$ can be obtained using the LIS system rather than point-to-point transmission. Finally, \cite{Basar_Frontier} suggests using LIS as an AP and compares the performance of the system. Furthermore, \cite{Basar_Frontier} shows that the LIS-AP system can provide ultra-reliable communication with an improvement of $1$ dB in the case of intelligent transmission.  Blind transmission in LIS-AP systems, however, yields the same performance of conventional-LIS systems.

\subsection{Phase Shift Error Effect on Transmission }
Adjusting the reflection phases, such that the signals at the destination combine coherently, enhances the communication performance. The calculation of accurate phase shifts, however,  is not feasible in practice. Hence, \cite{phase} studies the performance of the LIS in terms of signal transmission with phase error having a  generic distribution. The authors in \cite{phase} analyze the LIS performance for a limited number of reflectors with two types of errors, which are phase estimation error and quantization error. The signal at the receiver's side is expressed as follows \cite{phase},
\begin{equation}
  Y= N \sqrt{\gamma _{0}} HX + W,
\end{equation}
where  $\gamma_{0}$  is the average SNR, $X$ is the transmitted symbol,  $N$ is the number of reflectors,   and $ W \sim \mathcal{CN} \left( 0,1 \right)$  is the normalized receiver noise. The channel gain $H$ is represented as
\begin{equation}\label{eq: phaseH}
 H = \frac{1}{N} \sum _{i=1}^{N} {H_{i1}} {H_{i2}} e^{j\Omega_{i}} \in \mathbb{C},
\end{equation}
where $ H_{i1}$  and $ H_{i2} $  are the complex fading coefficients between source to reflector and reflector to destination, respectively.

To maximize the SNR, the practical phase $\Phi _{i}$  is set to cancel the summation of the phases  $H_{i1} $  and  $H_{i2}$  that denotes the overall phases.  Also, in (\ref{eq: phaseH}), $\Omega $ refers to the phase noise which has a normal distribution between $\left[ - \pi , \pi \right]$. The work in \cite{phase} further  assumes that $ \theta_{i},~ i=1, \ldots .,N $  are independent and identically distributed having a common characteristic function, which is  labeled as trigonometric/circular moments.
When $N$ is large, $H$ has a complex normal distribution with non-circular symmetry; however, \cite{phase} examines the performance when $N$ is finite.  When there is no phase error, which is the ideal case, the coefficient $H$ is real and  $H \sim \mathcal{N} \left( a^{2}, ( 1-a^{4}) /N \right)$, with $ a$ as the power parameter. When phase errors exist, $ H \sim \mathcal{CN} (0, \frac{1}{N})$, which indicates that there is a lack of information about $ H_{i1}$  and $H_{i2}$  phases.

The study in \cite{phase} shows that the communication channel via LIS with phase error is the same as a point-to-point channel with Nakagami fading, where the parameters of both are influenced by phase uncertainty via the first two circular moments. Moreover, the average SNR  increases with $N^2$, and the diversity order increases with $N$. Most importantly, when the number of reflectors is limited, the numerical analysis of error rate verifies that the LIS performance is vigorous despite the phase errors.

\subsection{Reflection Probability of LIS Systems}
The work in \cite{Renzo_Reflection} investigates the reflection probability of randomly distributed objects in the LIS-aided wireless networks, where the reflection probability is a function of the LIS length and the locations of the transmitter, the receiver, and the targeted object. For an object to be a reflector, two events must hold. First, the transmitter and the receiver have to be on one side of the reflector, lying on an infinite line that intersects the infinite line of the object segment, which we refer to as Event~1. Second, a perpendicular bisector line connecting the transmitter and the receiver must intersect the object segment itself, which we refer to as Event~2. Therefore, for LIS to reflect, the following probability must hold, 
\begin{equation}
     \text{Pr} \left( \text{LIS to reflect} \right) = \text{Pr} \{  \text{Event~1}  \cap~ \text{Event~2} \}.
\end{equation}
This study indicates that the reflection probability of objects coated with metasurfaces is independent of its length due to its capability of adjusting the reflection angles, covering more than what is expected by Snell's law \cite{Renzo_Reflection} \cite{relay}.  It is clear that for different length values, the reflection probability is almost constant, which indicates that small-sized objects can attain a high likelihood of being a reflector when they are placed appropriately. Therefore, it can reduce the cost of manufacturing and deployment over large-sized reflecting surfaces. Moreover,   large-scale  deployment of LIS improves the coverage area, reducing the blind-spots for terrestrial cellular BSs \cite{kishk2020exploiting}.
Nevertheless,  the authors in \cite{ozdogan2020} show that careful deployment  and proper  selection of phase shifts are necessary to get the full potential of LIS systems.

\subsection{Impact of Size on Performance of LIS Systems}
Recent theoretical works in \cite{relay} analyze the performance of different-sized LIS systems by comparing them with relay stations in terms of the average SNR as a function of the number of elements and the end-to-end transmission distance. Furthermore, \cite{relay} states that for LIS to be large and act as anomalous mirrors, the geometric size of each component of LIS has to be $10$ times greater than the radio wavelength of the impinging signal. Thus, whenever less than this threshold value, the intelligent surfaces are considered small and act as diffusers. The study in \cite{relay} highlights how the average SNR is scaled differently according to the type of connectivity with either relay stations, large intelligent surfaces, or small intelligent surfaces. For example, a relay station average SNR is scaled by a factor of

\begin{equation}\label{equ:ur1}
 N \text{min}  \left( d_{SR}^{-2},~d_{RD}^{-2} \right),
\end{equation}
a large intelligent surfaces system average SNR is scaled by a factor of 
\begin{equation}\label{equ:ur2}
    N^{2} \left( d_{SR}+ d_{RD} \right) ^{-2},
\end{equation}
 and a small intelligent surfaces system average SNR is scaled by a factor of 
\begin{equation}\label{equ:ur3}
     N^{2} \left( d_{SR}^{2}~d_{RD}^{2} \right) ^{-1},
\end{equation}
where $ d_{SR} $  refers to the distance between the source and relay/LIS,   $ d_{RD}$ refers to the distance between the relay/LIS and destination, and $N$ is the number of antenna elements at the relay station or the LIS.

In \cite{meta}, the authors show that in relay stations, the average end-to-end SNR grows linearly with $N$ because the total power remains constant since the power is distributed among all the antennas, as in  (\ref{equ:ur1}). The average SNR, however,  increases quadratically with $N$ in  large and small intelligent surfaces systems,  as in (\ref{equ:ur2}) and (\ref{equ:ur3}).  The study in \cite{arun2019rfocus} claims that this is especially the case because each element behaves as a separate reflecting mirror that scales the power by the transmittance before reflecting it.

Moreover, in relay stations, the SNR scales with the smallest distance between the two paths, as in (\ref{equ:ur1}). In LIS, the SNR scales with the total transmission distance, as in (\ref{equ:ur2}). In small intelligent surfaces, however,  the signal from each meta-element may combine, resulting in the scaled SNR, as in (\ref{equ:ur3}). Hence, it is undeniable that the LIS outperform the relay stations and small intelligent surfaces since the SNR in the case of LIS has the most significant scaling law that results in having a better transmission rate.  Furthermore, for specific values of $N$ at $2.6$ GHz and $28$ GHz as examples, LIS significantly double the rate of transmission \cite{relay}.

 \section{The Potential of Positioning and Coverage in LIS Systems}\label{sec: position}
The next-generation wireless communication networks anticipate enabling accurate location-based services where mmWaves and THz technologies will achieve a centimeter level of accuracy \cite{sarieddeen2019generation}. In this regard,   LIS can be deployed both for indoor and outdoor environments, making it one of the options for accurate positioning and localization. This section, therefore, surveys the recents works which study the potential of positioning using LIS systems.
Received signal strength (RSS) based positioning methods, in general, require high RSS values and coverage probability \cite{Khan2017}. In particular, \cite{CRLB} derives the  Cram\'{e}r-Rao  Bound for UE localization and positioning, whereas \cite{Hu_datatx} studies the performance of LIS for positioning and localization and comparing the accuracy of distributed and centralized LIS systems. On the other hand, \cite{positionmmWave} examines the potential of mmWave MIMO system positioning with and without the aid of an LIS system. Finally, \cite{spherical} expands the LIS model from planar to spherical surfaces and assesses its RSS and coverage.

  \subsection{Positioning in Centralized and Distributed LIS Systems}
In  \cite{Hu_datatx} and \cite{CRLB}, the authors derive the Cram\'{e}r-Rao Lower Bounds (CRLB) for UE navigation using the uplink signal of the LIS system.  Notably, the analysis of \cite{Hu_datatx} leverages the LIS system to provide robustness by subdividing a given surface area into smaller units of intelligent surfaces (i.e., distributed LIS system); however, the distributed LIS system may increase the complexity and feedback overheads. The work \cite{Hu_datatx} shows that the CRLB for a UE positioned at the perpendicular bisected line of the LIS decreases linearly as a function of the surface area. When a UE is not  positioned as such, there is not a closed-form solution for the CRLB; however, \cite{Hu_datatx} analytically approximates and shows that CRLB decreases in a quadratic fashion.

\begin{figure}
    \centering
    \begin{minipage}{0.45\textwidth}
        \centering
        \includegraphics[width=0.8\textwidth]{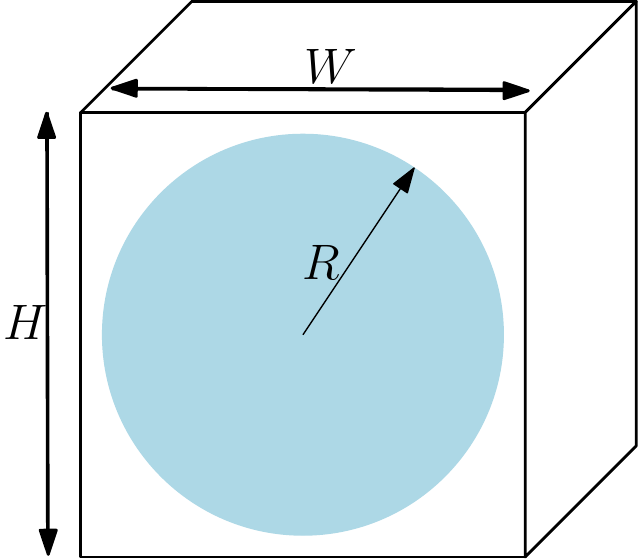}
        \caption{Centralized deployment of LIS system.\label{fig: cent}}
    \end{minipage}\hfill
    \begin{minipage}{0.45\textwidth}
        \centering
        \includegraphics[width=0.8\textwidth]{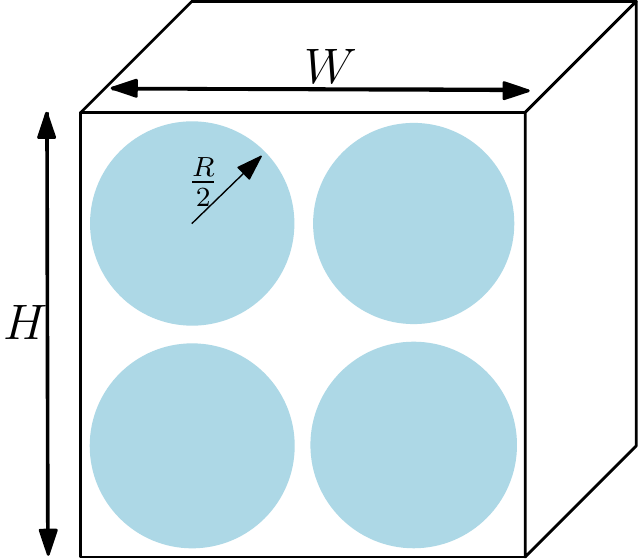} 
        \caption{Distributed deployment of LIS system scaled by half.\label{fig: dist}}
    \end{minipage}
\end{figure}

 In \cite{CRLB}, a comparison between the centralized (Fig. \ref{fig: cent}) and distributed (Fig. \ref{fig: dist}) deployment of LIS, in terms of coverage probability, is also provided.
The centralized LIS system refers to having each surface in the system as a whole unit, whereas the distributed LIS system refers to having each surface divided into several smaller independent, intelligent units. The study \cite{CRLB} further shows that the distributed implementation has lower CRLB for the $x-$ and $y-$ dimensions than the centralized implementation for the same total surface area. Also, the probability of coverage for a distributed LIS system is significantly better than the centralized one, which eventually improves the positioning performance. The study \cite{CRLB} also verifies that the distributed implantation facilitates a flexible deployment, yet requires specialized hardware for phase calibration and cooperation between the  LIS'  sub-units.
  Also, the distributed implementation allows the units to be replaced when needed without affecting the whole system. Nevertheless, hardware impairments are minimized in distributed LIS \cite{Capacity_hu}, \cite{Hu_datatx}.

 \subsection{Positioning in mmWave Aided Systems}
The quality of RSS values is a critical factor in determining the position of UEs. The study \cite{positionmmWave} investigates leveraging the phase and number of elements of the LIS system to aid the mmWave systems in obtaining high accuracy of positioning. The authors in \cite{positionmmWave} study the difference in the positioning performance of a conventional mmWave system and an LIS-assisted mmWave system. The study in \cite{positionmmWave} follows the same procedure of \cite{CRLB} to obtain the CRLB performance bound of both systems. Further, \cite{positionmmWave} controls the phase and amplitude of the propagated waves by using LIS, where the study ultimate goal is to use the LIS-aided system to minimize the error in the position estimation. The results of \cite{positionmmWave} show that an LIS-aided mmWave MIMO system has a better positioning performance and smaller orientation error bound, when increasing the number of elements, even for as little as 40 elements.
\subsection{Positioning Using Spherical LIS}
 The authors in \cite{spherical} expand the system of \cite{Hu_datatx} to a spherical LIS instead of a planar one. Using spherical surfaces instead of planar has many advantages; for instance, it can act as a reflecting surface as well as a relaying surface. One part of the spherical LIS can work as a reflecting surface, while the other part can act for relaying the signals to UEs. The latter is particularly useful when a UE is blocked from its serving BS. Also, unlike traditional planar LIS,  a rotating user over spherical LIS would not impose changes in the information-theoretical properties, which yields an RSS gain for a mobile UE. Hence, \cite{spherical} derives the CRLB for RSS based spherical LIS systems in terms of the normalized distance given by $\nu = r/\cos{\theta}$, where $r$ is the radius for the sphere and $\theta$ is the elevation angle of the UE. The study, then, compares the RSS and CRLB of a spherical LIS system with a planar LIS system of the same surface area, i.e., the radius of the disk-shaped planar LIS is $r_p = \sqrt{2}r$. The CRLB of spherical LIS is derived by evaluating the Fisher-information matrix and uses RSS to compute CRLB in terms of the sphere radius. The CRLB of spherical LIS, therefore, is given by \cite{spherical}
 \begin{equation}\label{eq: crlbsp}
     \text{CRLB}_{sph} =  4 \nu^4(\nu^2 -1),
 \end{equation}
 while the CRLB of planar disk shaped LIS is
 \begin{equation}\label{eq: crlbpl}
    \text{CRLB}_{pl} =   4 (\nu^2 + 1)^3.
 \end{equation}
It can be seen from (\ref{eq: crlbsp}) and (\ref{eq: crlbpl}) that the CRLB of spherical LIS is smaller than the planar LIS, which makes it more accurate for positioning of the UEs. The results of \cite{spherical} verify the above findings, and show that especially when $r$ increases, the spherical LIS outperforms the traditional planar one.

\section{Open Research Issues}\label{sec:openresearch}
Since the beginning of the recent active LIS research era, there has been a plethora of studies that focused on the joint beamforming optimization problems, theoretical SNR and SEP derivations, channel estimation, and SINR maximization. Furthermore, many researchers investigated the application of machine learning tools and the evaluation of LIS potential for the mmWave/THz, free-space optics, and visible light applications \cite{basar2019wireless}.
For example, in \cite{Beam_Wu}, the authors explored the problem of joint active and passive beamforming design that minimizes the total transmit power at the BS. They employed optimization techniques, such as semidefinite relaxation and alternating optimization, to solve the non-convex optimization problem. For SEP derivations, the author in \cite{Basar_Frontier} proposed a SEP mathematical framework for LIS systems by studying the ability of using an LIS as an AP transmitter.
However, in spite of the timely studies mentioned above, there are open research issues that should be tackled to guarantee a high level of reliability in intelligent wireless networks. In the following, we explore few of the promising upcoming research challenges.

\subsection{Realistic Optimization Frameworks}
Classical optimization problems need to be addressed and reformulated due to the fundamental challenges that IRS implementation faces. Such issues are due to the fact the most of the current works are rather based on non-practical assumptions, e.g., perfect channel estimation (i.e., perfect CSI assumptions), ignoring internal losses and far-field radiations, accurate beamforming and beamstearing, single-antenna UE, optimal precoding, etc. Hence, future works need to revisit such assumptions and examine the reliability of IRSs through realistic approaches.

Moreover, while the majority of the studied optimization problems in LIS scenarios focus on maximizing the EE, throughput, and SINR, several objectives have not been addressed in the literature yet. Firstly, the literature lacks undertaking LIS when it comes to ultra-reliable low latency communication. To this end, we propose minimizing the overall power consumption of LIS users subject to transmit power budget and reliability constraints in terms of probabilistic queuing delay. In particular, the reliability measure should account for events where users' queue length exceeds a certain threshold. For instance, after accounting for unknown CSI and networks dynamics, the problem can be solved using federated learning, i.e., through distributed approaches. Leveraging such approaches would enable multiple learners to define a set of local parameters from the existing training data where they can share their local models rather than sharing the training data.

Secondly, one can look into the energy harvesting (EH) aspect in LIS systems, whereby a possible source of energy can come from the ubiquitous radio transmitters. EH provides green and sustainable solutions to the power loss. Hence, EH-empowered LIS system can work as an energy harvester, where it converts the incident signal into electrical energy \cite{sun2020reconfigurable, Zhao2020}. Nonetheless, RF sources suffer from low incident power levels, mainly depending on the transmitted wave frequency, the antenna gain, and the communication range. Therefore, we propose maximizing the receiver incident power subject to the transmitter power and the gap between the transmitter and the LIS. Adjusting the position angles of the antennas can maximize the incident power of the received signal, where the optimal position angles can be iteratively computed using the adaptive gradient ascent method.

Finally, several IRS-assisted communication systems works such as \cite{Wu_BeamOpt}, \cite{Weighted}, and \cite{Beam_Wu} utilized alternating optimization-based algorithms to tackle the optimization problem. Applying such a technique can jeopardize the process of reaching a joint optimal solution. Hence, alternatingly solving the subproblems is costly due to the high computational complexity. Therefore, developing advanced algorithms that can eliminate the use of alternating optimization can be a promising research topic in this direction. One possible approach that future studies must exploit is data-driven optimization. IRS systems are complex to analyze and design compared to conventional wireless networks. Therefore, to reduce the complexity system, effective data-driven optimization techniques that depend on deep learning, transfer learning, and reinforcement learning should be considered. In short, machine learning can prove to provide a powerful approach to improve the IRS-based communication systems performance. 


Another promising future research direction is to envision an optimized system that aims at optimizing the fraction overlap between the uplink and downlink, i.e., introducing a full-duplex IRS-assisted communication system \cite{xu2020resource, pan2020fullduplex}. Incorporating the IRS's full-duplex characteristics can further optimize the communication systems in terms of power allocation and beamforming. Besides full-duplex communications, integrating cooperative NOMA and IRS is an exciting future research direction, through accounting for the proper constraints and objectives when designing such systems.

\begin{figure}
    \centering
    \includegraphics[width = \linewidth]{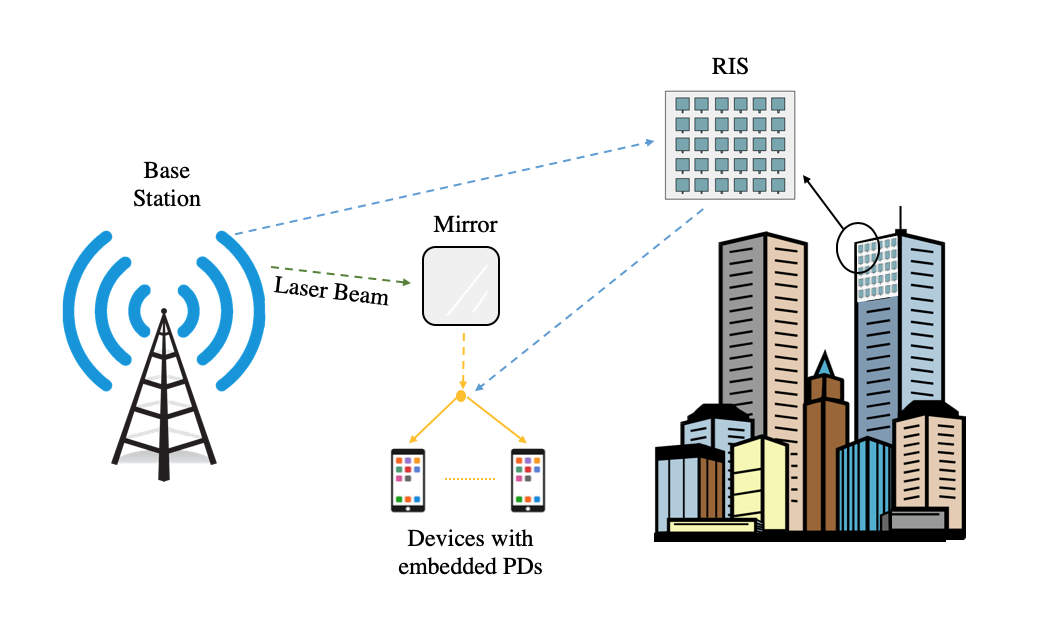}
    \caption{RF-VLC hybrid system for RISs. \label{fig:my_label}}
    
\end{figure}

\subsection{Hybrid Systems: RF-VLC}
RIS can empower the practical implementation of beyond 5G (B5G) systems by means of controlling the randomness of the propagation environment. RIS provides other advantages, including EE and full-band response. Nevertheless, B5G networks require significant enhancements in mobile broadband, enabling ultra-reliable low-latency communications \cite{basar2019wireless}. To this end, deploying hybrid systems can help in providing fast, efficient, and reliable communication networks. Specifically, visible light communication (VLC) has been a prominent research area in advanced communication systems \cite{vlc, Amr2020}. VLC has the potential of providing ultra-high bandwidth, robustness to electromagnetic interference, and inherent physical security. Both RIS-RF and VLC can be used in an outdoor environment as shown in Fig.~\ref{fig:my_label}, where mirrors reflect visible light (VL) signals to a photodetector in the users' devices while RISs are used to reflect RF signals from the BS. VL provides safe and health-friendly communication schemes that can be exploited in health-constrained centers such as hospitals. Furthermore, RIS can be used to complement VLC whenever LoS conditions fail, thus, supporting future wireless networks. Deploying this hybrid system model would provide a reliable communication scheme that compensates for potential failure of one of the connecting links. At the moment, most of the analysis in LIS exists only for RF and mmWave communications. Hence, future research can further investigate the performance of hybrid technologies in the LIS realm.

\subsection{Coating EM Materials}

Recall from Sec. I that RIS consist of controllable EM materials, where the antennas are coated with reconfigurable thin layers of EM materials to control the propagation of signals. Therefore, tunable materials should be used to adjust the signals' phase shifts, thereby adapting the transmitted signals according to the changes in the wireless environment \cite{basar2019wireless}. Meta-surfaces are the key enablers of such technologies \cite{meta}. Moreover, a widely tunable bandgap material is suggested to support full-band response, where theoretically, it can operate at at any frequency
band. To this end, to best realize the RIS systems, the usage of graphene is proposed as a future research direction. The graphene bandgap ranges from $0$ eV to $4.66$ eV. Consequently, it operates from the radio wave band to the infrared band. Graphene reconfigurable meta-surfaces can further achieve beam steering, beam focusing, and wave vorticity control by means of local tuning. For instance, phase control in graphene metasurfaces is achieved by changing its conductivity via electrostatic biasing. Due to such unique characteristics, future research directions should exploit how to adjust the performance of graphene-based RIS systems through optimizing the system conductivity via electrostatic biasing.

\subsection{Health Issues}
RF technologies are proliferating with the emergence of 6G. According to \cite{Zanaty2020survey} and \cite{RF}, a wide range of human health concerns are correlated with exposure to the RF radiations.
The associated health issues with EM radiation exposure have been an open research topic for decades. However, the recent advent of the LIS in indoor communication scenarios opened the door for more concerns regarding the possible health risks. Mostly, indoor mmWave environments are studied for future applications as they offer large bandwidth for enabling high data rates. However, unlike cellular phone frequencies, mmWave radiations are high-frequency signals that have relatively deep penetration in the human body. Hence, the primary concern is the heating of the skin and eyes resulted from the body absorption of mmWaves. In reality, recent studies show that current estimating power density methods are not reliable for determining the exposure compliance at close mmWave interaction \cite{admin_2019}. Therefore, to maintain an efficient high-rate LIS system, we propose considering an optimization problem that maximizes the data-rate, subject to health constraints, by means of adjusting the distance between several LISs, particularly in indoor environments. According to the National Toxicology Program (NTP) studies, small rooms with passive elements placed in close proximity to reflect and generate EM radiation caused health issues to the exposed rats in the long term  \cite{Zanaty2020survey}\cite{national2018toxicology}. While increasing the distance would decrease the RF exposure of users, it would affect the system overall performance. Future research studies must, therefore, consider such trade-offs in performance versus health issues by properly adjusting different network parameters.

\subsection{Integration of 5G and 6G Technologies}
LIS is one of the revolutionary and potential physical layer technologies that generate a new communication paradigm that meets the requirements of future 6G networks\cite{yuan2019potential,    Basar_Frontier}. Also, smart radio environments may have a potential impact on the upcoming 6G technology markets, facilitating substantive improvements in spectral efficiency with cost-effective solutions. One of the undiscovered research directions is the integration of intelligent surfaces with emerging 5G technologies, such as IoT \cite{zhang2020large}, drones-aided communications \cite{Ge2020}, beamforming \cite{Nadeem2020}, and physical layer security \cite{Guan2020}. For 6G technology, an LIS system has the potential to provide a pervasive and reliable wireless communication service, while suppressing additional interference components such as noise and inter-user interface through both NLOS and LOS paths. Since the LISs decrease network interference level, they are expected to improve the network capacity and user performance in 6G networks, especially for high-density user environments such as airports and stadiums. Even though the unique benefits of LIS help creating a favorable wireless communication channel, its use-cases and application scenarios that meet new user requirements and networking trends of 6G technology are still in their exploratory phase. Furthermore, the economic impact and its sustainability of LIS-assisted smart radio environment on B5G markets are key research questions that need to be further studied \cite {renzo2019smart,    Basar_Frontier}.

\subsection{Localization Using LIS Systems}
Positioning using LIS-aided mmWave systems can be analyzed and studied by jointly considering the design of mmWave beamformer and LIS phase shifters. Also, the localization performance of distributed spherical LISs and centralized spherical LIS systems can be investigated. Moreover,  the study of the CRLB of LIS systems in the presence of NLoS channels can be further pursued in future works.  Also, the performance regarding the positioning accuracy of LIS-aided systems highly depends on the placement of the reflectors and metasurfaces. Therefore, it is important to find the optimal locations to place these reflectors and metasurfaces. The placement of these materials is a challenging task, which is an inverse problem of channel modeling. In channel modeling, being aware of the deployment, we can obtain the channel state information using various channel modeling techniques such as ray tracing, etc. While based on the desired channel state information, the optimal deployment of the materials in LIS is an inverse task, the solution of which remains an open problem.

\section{Conclusion}\label{sec:conc}
Large intelligent surfaces (LIS) are a promising physical layer technology for B5G systems. Such technology does not only enhance wireless systems QoS, but also reduces the large power consumption as compared to traditional networks. LIS are made up of re-configurable EM meta-materials that are capable of modulating data onto the received signals, customizing changes to the radio waves, and intelligently sensing the environment. This paper provides a unique blend that surveys the principles of physical operation of LIS, together with their optimization and performance analysis frameworks. The paper first introduces the LIS technology and its working principle. Then, it presents various optimization techniques that aim to optimize specific objectives, namely, maximizing energy efficiency, power, sum-rate, secrecy-rate, and coverage. The paper afterwards discusses various relevant performance analysis works including capacity analysis, the impact of hardware impairments on capacity, uplink/downlink data rate analysis, and outage probability. The paper further presents the impact of adopting the LIS technology for positioning applications. Finally, we identify numerous exciting open challenges for LIS-aided B5G wireless networks, including new resource allocation problems, hybrid RF-VLC systems, health considerations, and localization. To the best of the authors' knowledge, this survey is the first of its kind which combines the technical aspects of mathematical optimization and performance analysis of LIS systems, and sheds light on promising research directions towards the formulations of practical problems in future B5G systems.

\bibliographystyle{IEEEtran}
\bibliography{IEEEabrv,bibliography}

\end{document}